\documentclass{sig-alternate-10pt}
\usepackage{etoolbox}

\usepackage{listings}
\usepackage{fancyhdr}
\usepackage{fancyvrb}
\usepackage{graphicx}
\usepackage{tabularx}
\usepackage{amsmath}
\usepackage{amssymb}
\usepackage{verbatim}
\usepackage{color}
\usepackage{caption}
\DeclareCaptionType{copyrightbox}
\usepackage{subcaption}
\usepackage{xspace}
\usepackage{enumitem}
\usepackage{helvet}
\usepackage{mathptmx}
\usepackage{lastpage}
\usepackage{datetime}
\usepackage{xfrac}
\usepackage{multirow}

\expandafter\let\csname ver@natbib.sty\endcsname\relax
\usepackage[maxnames=100,sortcites]{biblatex}
\makeatletter
\newcommand\subparagraph{%
  \@startsection{subparagraph}{5}
  {\parindent}
  {3.25ex \@plus 1ex \@minus .2ex}
  {-1em}
  {\normalfont\normalsize\bfseries}}
\makeatother
\usepackage{titlesec}
\let\subparagraph\relax 

\titlespacing*\section{0pt}{6pt plus 2pt minus 2pt}{2pt plus 2pt minus 2pt}
\titlespacing*\subsection{0pt}{6pt plus 2pt minus 2pt}{2pt plus 2pt minus 2pt}

\newcommand{\toappear}[1]{}

\definecolor{orange}{rgb}{1,0.5,0}
\definecolor{darkslateblue}{rgb}{0.28,0.24,0.55}
\definecolor{gy}{rgb}{0,1,0.8}

\newcommand{\cam}[1]{{\color{blue}#1}}
\newcommand{\extended}[1]{{\color{red}#1}}
\newcommand{\nextended}[1]{}

\newenvironment{ecam}{\color{blue}}{}
\newenvironment{eext}{\color{red}}{}

\renewenvironment{ecam}{}{}
\renewenvironment{eext}{}{}
\renewcommand{\cam}[1]{#1}
\renewcommand{\extended}[1]{#1}
\renewcommand{\nextended}[1]{}

\newcommand{\smallsec}[1]{\medskip\noindent{\bf #1}}
\newcommand{\todo}[1]{}
\newcommand{\note}[1]{{\color{red} #1}}
\newcommand{\ignore}[1]{}
\newif\ifcpintegration
\cpintegrationfalse

\begin{document}
\begin{sloppypar}
\title{Millions of Little Minions: Using Packets for Low Latency\\Network Programming and Visibility\\(Extended Version)}

\conferenceinfo{SIGCOMM}{'14, Chicago, USA}



\author{\sffamily Vimalkumar Jeyakumar$^1$, Mohammad Alizadeh$^2$, Yilong Geng$^1$, Changhoon Kim$^3$, David Mazi\`eres$^1$\\
\normalsize \texttt{jvimal@cs.stanford.edu,\{alizade,gengyl08\}@stanford.edu, chkim@barefootnetworks.com,}\\\url{http://www.scs.stanford.edu/~dm/addr}\\
\begin{tabular}{ccc}
\sffamily $^1$Stanford University & \sffamily $^2$Cisco Systems & \sffamily $^3$Barefoot Networks \\
\sffamily Stanford, CA, USA  & \sffamily San Jose, CA, USA  & \sffamily Palo Alto, CA, USA
\end{tabular}}
\maketitle


\abstract{

This paper presents a practical approach to rapidly introducing new
dataplane functionality into networks: End-hosts embed tiny programs
into packets to actively query and manipulate a network's internal
state.  We show how this ``tiny packet program'' (TPP) interface gives
end-hosts unprecedented visibility into network behavior, enabling
them to work with the network to achieve a desired functionality.  Our design
leverages what each component does best: (a) switches forward and execute
tiny packet programs (at most 5~instructions) in-band at line rate, and (b)
end-hosts perform arbitrary (and easily updated) computation on network
state.  By implementing
three different research proposals, we show that TPPs are
\emph{useful}.  Using a hardware prototype on a NetFPGA, we show our
design is \emph{feasible} at a reasonable cost.

}

\section{Introduction}
\label{sec:introduction}
Consider a large datacenter network with thousands of switches.
Applications complain about poor performance due to high flow
completion times for a small subset of their flows.  As an operator,
you realize this symptom could be due to congestion, either from
competing cross traffic or poor routing decisions, or alternatively
could be due to packet drops at failed links.  In any case, your goal
is to diagnose this issue quickly.  Unfortunately, the extensive use
of multipath routing in today's networks means one often cannot
determine the exact path taken by every packet; hence it is quite
difficult to triangulate problems to a single switch.  Making matters
worse, if congestion is intermittent, counters within the network will
look ``normal'' at timescales of minutes or even seconds.

Such issues would be straightforward to debug if one could examine
relevant network state such as switch ID, queue occupancy, input/output
ports, port utilization, and matched forwarding rules
\emph{at the exact time each packet was forwarded}, so as
to reconstruct what exactly transpired in the dataplane.  In the above
example, end-hosts could use state obtained from millions of
successfully delivered packets to explicitly pinpoint network links
that have high queue occupancy (for congestion), or use switch and
port IDs to verify that packets were correctly routed, or use path
information to triangulate network links that cause packet drops due
to link failures.  In short, the ability to correlated network state
to specific packets would be invaluable.

Can packets be instrumented to access and report on switch state?  To
date such state has been locked inside switches.  This paper describes
a \emph{simple, programmable} interface that enables end-hosts to
query switch memory (counters, forwarding table entries, etc.)\ from
packets, directly in the dataplane.  Specifically, a subset of packets
carry in their header a tiny packet program (TPP), which consists of a
few instructions that read, write, or perform simple, protocol-agnostic
computation using switch memory.

A key observation in this paper is that having such programmable
and \emph{fast} access to network state benefits a broad class of
network tasks---congestion control, measurement, troubleshooting, and
verification---which we call \emph{dataplane} tasks.  We show how the
TPP interface enables end-hosts to rapidly deploy new functionality by
refactoring many network tasks into: (a) simple TPPs that execute on
switches, and (b) expressive programs at end-hosts.

TPPs contrast to three approaches to introduce new dataplane
functionality: (1) build custom hardware for each task, (2) build switches
that can execute arbitrary
code~\cite{tennenhouse2002towards,schwartz1999smart}, or (3) use FPGAs
and network processors~\cite{lu2011serverswitch}.  Each approach has its own drawbacks:
Introducing new switch functionality can take many years; switch hardware has
stringent performance requirements and cannot incur the penalty of
executing arbitrary code; and FPGAs and network processors are simply
too expensive at large scale~\cite{glen2013hardware}.  \cam{Instead, we
argue that if we could build new hardware to support just
\emph{one} simple interface such as the TPP, we can leverage end-hosts to
implement \emph{many} complex tasks at software-development timescales.}

TPPs can be viewed as a particular, reasoned point within
the spectrum of ideas in Active
Networking~\cite{tennenhouse2002towards,schwartz1999smart}.  In many Active
Networks formulations, routers execute arbitrary programs that actively control
network behavior such as routing, packet compression, and
\mbox{(de-)}duplication.  By contrast, TPP instructions are so simple
they execute within the time to forward packets at line-rate.  Just a
handful of TPP instructions, shown in Table~\ref{tab:instructions},
providing access to the statistics in Table~\ref{tab:stats}, proved
sufficient to implement several previous research proposals.

\begin{table}
\centering\small
\begin{tabular}[t]{|p{0.1\textwidth}|p{0.32\textwidth}|}\hline
{\bf Instruction}  &  {\bf Meaning} \\\hline
{\tt LOAD, PUSH} & Copy values from switch to packet \\\hline
{\tt STORE, POP} & Copy values from packet to switch \\\hline
{\tt CSTORE} & Conditionally store and execute subsequent operations \\\hline
{\tt CEXEC} & Conditionally execute the subsequent instructions \\\hline
\end{tabular}\caption{The tasks we present in the paper require
support only for the above instructions, whose operands will be clear
when we discuss examples.  \cam{Write instructions may be selectively
disabled by the administrator.}}\label{tab:instructions}\vspace{-1em}
\end{table}

\subsection{Goals}
Our main goal is to expose network state to end-hosts through the
dataplane.  To benefit dataplane tasks, any interface should satisfy
the following requirements:
\begin{itemize}[noitemsep,leftmargin=1em,noitemsep]
\item {\bf Speed}: A recent study shows evidence that switch
CPUs are not powerful and are unable to handle more than a few hundred
OpenFlow control messages/second~\cite{huang2013high}.  Our experience
is that such limitations stand in the way of a whole class of
dataplane tasks as they operate at packet and round-trip timescales.
\item {\bf Packet-level consistency}:  Switch state such as link queue
occupancy and forwarding tables varies over time.  Today, we lack any
means of obtaining a consistent view of such state as it pertains to
each packet traveling through the network.
\item {\bf Minimality and power}:  To be worth the effort,
any hardware design should be simple, be sufficiently expressive to
enable a diverse class of useful tasks, and incur low-enough overhead
to work at line rates.
\end{itemize}

This paper presents a specific TPP interface whose design is largely
guided by the above requirements.


\smallsec{Non-Goals}: It is worth noting that our goal is not to be flexible
enough to implement any, and all dataplane network tasks.  For
instance, TPPs are not expressive enough to implement per-packet scheduling.
Moreover, our design is
for networks owned and operated by a single administrative entity
(e.g., privately owned WANs and datacenters).  We do not advocate
exposing network state to untrusted end-hosts connected to the
network, but we describe mechanisms to avoid executing untrusted
TPPs (\S\ref{subsec:security}).  Finally, a detailed design for
inter-operability across devices
from multiple vendors is beyond the scope of this paper, though we
discuss one plausible approach (\S\ref{sec:discussion}).

\subsection{Summary of Results}
Through both a software implementation and a NetFPGA prototype, this
paper demonstrates that TPPs are both useful and feasible at line
rate.  Moreover, an analysis using recent data~\cite{glen2013hardware}
suggests that TPP support within switch hardware can be realized at an
acceptable cost.

\smallsec{Applications}:  We show the benefits of TPP by refactoring
many recent research proposals using the TPP interface.  These tasks
broadly fall under the following three categories:

\begin{itemize}[noitemsep,leftmargin=1em]
\item {\bf Congestion Control}: We show how end-hosts, by periodically 
querying network link utilization and queue sizes with TPP, can
implement a rate-based congestion control algorithm (RCP) providing
max-min fairness across flows.  We furthermore show how the TPP
interface enables fairness metrics beyond the max-min fairness for
which RCP was originally designed~(\S\ref{subsec:rcp}).

\item {\bf Network Troubleshooting}: TPPs give end-hosts detailed
per-packet visibility into network state that can be used to implement a recently
proposed troubleshooting platform called NetSight~\cite{netsight}.  In
particular, we walk through implementing and deploying {\tt ndb}, a
generalization of traceroute introduced by
NetSight~(\S\ref{subsec:netsight}).
\extended{
\item {\bf Network Monitoring}: TPPs can be used in a straightforward way to do
network monitoring, but we also show how to refactor new kinds of
measurement tasks: For example, OpenSketch~\cite{yu2013software}
proposed switch modifications to increase accuracy of five different
measurement tasks while incurring low storage overhead.  We show how to
achieve similar functionality using network visibility offered by TPPs.
In particular, we walk through
using TPPs to count the number of unique source IP addresses that
communicate over all network links in the core of the
network~(\S\ref{subsec:sketch}).}
\cam{
\item {\bf Network Control}: We also demonstrate how low-latency visibility offered
by TPPs enables end-hosts to control how traffic is load balanced across
network paths.  We refactor CONGA~\cite{conga}, an in-network load-balancing
mechanism implemented in Cisco's new ASICs, between end-hosts and a network
that supports only the TPP interface.}
\end{itemize}

\begin{table}
\centering\small
\begin{tabular}[t]{|l|p{0.35\textwidth}|}\hline
{\bf Statistics}  &  {\bf Examples} \\\hline
Per-Switch  &
Switch ID, counters associated with the global L2 or L3 flow tables, flow table version number, timestamp. \\\hline
Per-Port  &
Link utilization, bytes received, bytes dropped, bytes enqueued, application-specific registers. \\\hline
Per-Queue &
Bytes enqueued, bytes dropped. \\\hline
Per-Packet &
Packet's input/output port, queue, matched flow entry, alternate routes for a packet.\\\hline
\end{tabular}
\caption{A non-exhaustive list of statistics stored in switches memory that
  TPPs can access \cam{when mapped to known memory locations.}  Many statistics are already tracked
  today but others, such as flow table version will have to be
  implemented.  Some statistics are read-only (e.g. matched flow
  entry, bytes received), but others can be modified (e.g. packet's
  output port).  \cam{See OpenFlow 1.4 specification~\cite[Table~5]{of1.4}
  for a detailed list of available statistics.}}\vspace{-1em}
\label{tab:stats}
\end{table}

\vspace{-1em}\smallsec{Hardware}:  To evaluate the feasibility of building a
TPP-capable switch, we synthesized and built a four-port NetFPGA
router (at 160MHz) with full TPP support, capable of switching minimum
sized packets on each interface at 10Gb/s.  We show the hardware and
latency costs of adding TPP support are minimal on NetFPGA, and
argue the same would hold of a real switch~(\S\ref{sec:evaluation}).
We find that the key to
achieving high performance is restricting TPPs to a handful of
instructions per packet (say five), as it ensures that any TPP
executes within a fraction of the its transmission time.

\smallsec{Software}:  We also implemented the TPP interface
in Open~vSwitch~\cite{pfaff2009extending}, which we use to demonstrate
research proposals and examples.  Additionally, we present a software
stack~(\S\ref{sec:endhost}) that enforces security and access
control, handles TPP composition, and has a library of useful
primitives to ease the path to deploying TPP applications.




\cam{
The software and hardware implementations of TPP,
scripts to run experiments and plots in this paper,
and an extended version of this paper describing more
TPP applications are all available online at \url{http://jvimal.github.io/tpp}.
}


\ignore{\section{Motivation for TPPs}
\label{sec:motivation}
Consider a large network with 100s of switches, in which one switch
has an \emph{intermittent} faulty line-card -- it works for 2~minutes,
and then fails to forward any packets for a minute.  Due to multipath routing, this
line-card affects traffic from a large fraction of users for a small
interval of time, before it is automatically reset.  Since the network
is large, control plane logs are too big to analyze.  Moreover, since
the line-card is functioning intermittently, counters look ``normal'' at
minute timescales, and do not record packet drops when the line-card
fails.  Such problems are not uncommon at scale: This problem took
3\sfrac{1}{2}~days to debug in a real cloud provider's network with
today's monitoring mechanisms.

A straightforward approach to debug such an issue would be to
instrument a large fraction of traffic (with low-overhead), that
continously reads statistics such as queue sizes, switch ID, input and
output ports, to \emph{reconstruct} what exactly transpired in the
dataplane.  Packets embedded with TPPs would be routed normally, and
after the packet reaches its destination, the statistics within the
packet can be analyzed: For example, in the above scenario, if the
analysis shows high loss rates despite no congestion (i.e. small queue
sizes), it is a useful red-flag worth probing further; the topology
information (switch and link IDs) within the packet helps narrow down
the search to a specific line-card.

Our (simple) observation is that having programmable access to network
state benefits a much broader class of network tasks: congestion
control, measurement, troubleshooting, and verification.  TPPs are a
generalization of this explicit approach: end-hosts access dataplane
statistics using packets.  This approach addresses the two drawbacks
of today's monitoring mechanisms: (i)~speed: end-hosts can obtain
statistics at round-trip timescales, and (ii)~consistent view: the
statistics are tied to the packet and provide useful context that is
otherwise hard to obtain in the control plane.

In the next section, we show several examples of network tasks that
require specialized hardware, but become feasible with minimal TPP
support from the ASIC.
}

\section{Example Programs}
\label{sec:examples}
\cam{We start our discussion using examples of dataplane tasks that can be implemented
using TPPs, showcasing the utility of exposing network
state to end-hosts directly in the dataplane.  Each of
these tasks typically requires new task-specific hardware changes;
however, we show how each task can be refactored such that the network only implements
TPPs, while delegating
complex task-specific functionality to end-hosts.  We will
discuss the following tasks: (i) micro-burst detection, (ii) a
rate-based congestion control
algorithm, (iii) a network troubleshooting platform, (iv) a congestion
aware, distributed, network load balancer}\extended{, and (v)
a low-overhead network measurement platform}.

\begin{figure*}[htb]
\centering
\begin{subfigure}[b]{0.40\textwidth}
\includegraphics[width=\textwidth]{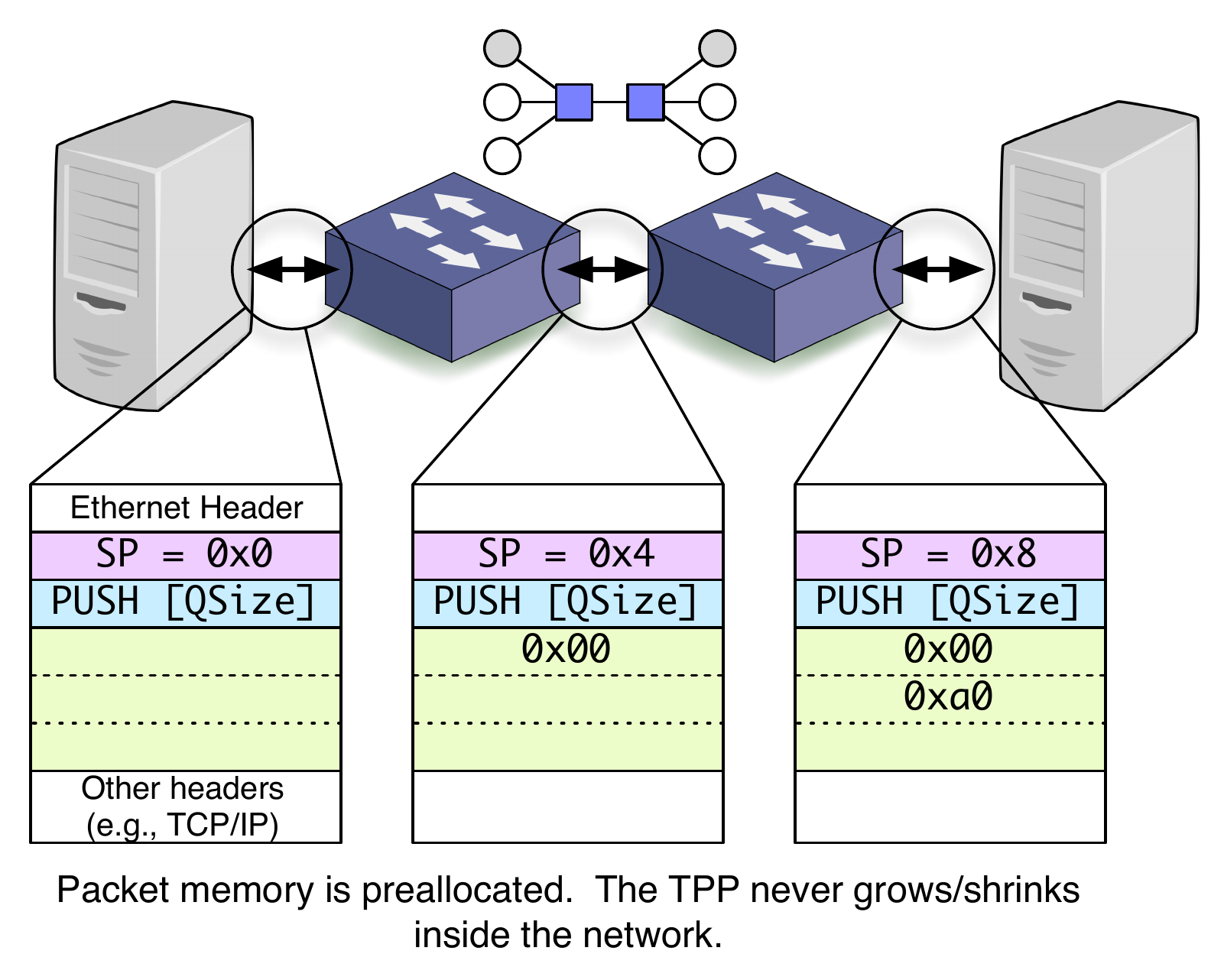}
\caption{Visualizing the execution of a TPP as it is
  routed through the network.}\label{fig:tpp-qsize-example-topo}
\end{subfigure}%
\quad
\begin{subfigure}[b]{0.40\textwidth}
\includegraphics[width=\textwidth]{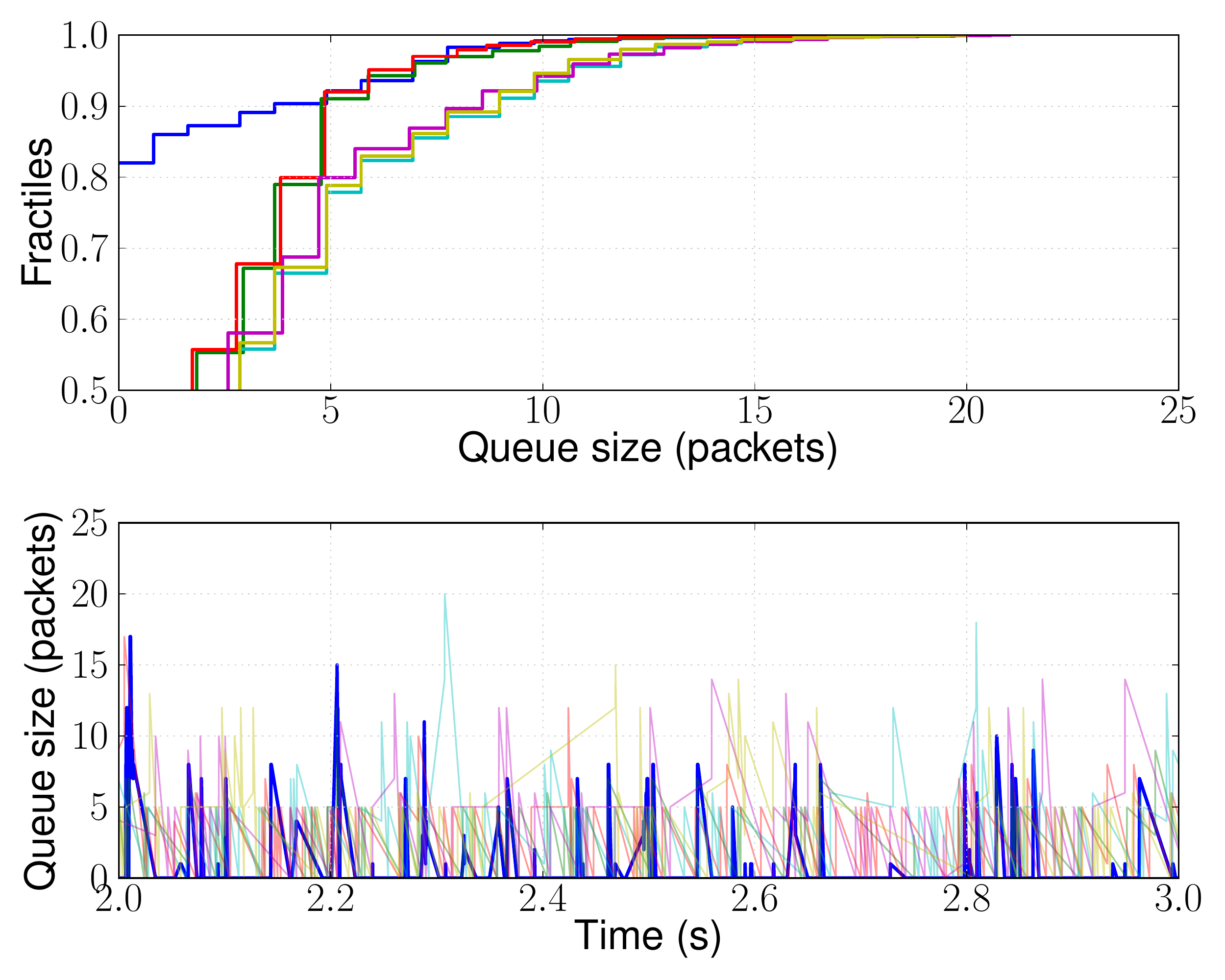}
\caption{CDF and time series of queue occupancy on 6 queues in
  the network, obtained from \emph{every} packet arriving at one
  host.}\label{fig:tpp-qsize-example-queues}
\end{subfigure}
\caption{TPPs enable end-hosts to measure queue occupancy evolution at
  a packet granularity allowing them to detect micro-bursts, which are
  the spikes in the time series of queue occupancy (bottom of Figure~\ref{fig:tpp-qsize-example-queues}).
  Notice from the CDF (top) that one of the queues is empty for 80\% of the time instants when
  packet arrives to the queue; a sampling method is likely to miss the bursts.}
\label{fig:tpp-qsize-example}
\end{figure*}


\smallsec{What is a TPP?}  A TPP is any Ethernet packet with a
uniquely identifiable header that contains instructions, some
additional space (packet memory), and an optional encapsulated
Ethernet payload (e.g.\ IP packet).  The TPP exclusively owns its
packet memory, but also has access to shared memory on the switch (its
SRAM and internal registers) through addresses.  TPPs execute directly
in the dataplane \cam{at every hop}, and are forwarded just like other
packets.  TPPs use a very minimal instruction set listed in
Table~\ref{tab:instructions}, and we refer the reader to
Section~\ref{sec:design} \cam{to understand the space overheads.
We abuse terminology, and use TPPs to refer both to the
programs and the packets that carry them}.

We write TPPs in a pseudo-assembly-language with a segmented address
space naming various registers, switch RAM, and packet memory.
We write addresses
using human-readable labels, such as \texttt{[Namespace:Statistic]}
or \texttt{[Queue:QueueOccupancy]}.  \cam{We posit that these addresses
be known upfront at compile time.}  For example, the mnemonic {\tt [Queue:QueueOccupancy]}
could be refer to an address \texttt{0xb000} that
stores the occupancy of a packet's output queue at each switch.



\subsection{Micro-burst Detection}
Consider the problem of monitoring link queue occupancy within the
network to diagnose short-lived congestion events (or
``micro-bursts''), which directly quantifies the impact of incast.  In
low-latency networks such as datacenters, queue occupancy changes
rapidly at timescales of a few RTTs.  Thus, observing and controlling
such bursty traffic requires visibility at timescales orders of
magnitude faster than the mechanisms such as SNMP or embedded web
servers that we have today, which operate at tens of seconds at best.
Moreover, even if the monitoring mechanism is fast, it is not clear
which queues to monitor, as (i) the underlying routing could change,
and (ii) switch hash functions that affect multipath routing are often
proprietary and unknown.

TPPs can provide fine-grained per-RTT, or even per-packet visibility
into queue evolution inside the network.  Today, switches
already track per-port, per-queue occupancy for memory management.  The
instruction {\tt PUSH [Queue:QueueOccupancy]} could be used to copy the queue
register onto the packet.  As the packet traverses each hop, the
packet memory has snapshots of queue sizes at each hop.  The queue
sizes are useful in diagnosing micro-bursts, as they are not an
average value.  They are recorded when the packet traverses the
switch.  Figure~\ref{fig:tpp-qsize-example-topo} shows the state of a
sample packet as it traverses a network.  In the figure, {\tt SP} is
the stack pointer which points to the next offset inside the packet
memory where new values may be pushed.  Since the maximum number of
hops is small within a datacenter (typically 5--7), the end-host
preallocates enough packet memory to store queue sizes.  Moreover, the
end-host knows exactly how to interpret values in the packet to obtain
a detailed breakdown of queueing latencies on all network hops.

This example illustrates how a low-latency,
programmatic interface to access dataplane state can be used by
software at end-hosts to measure dataplane behavior that is hard to
observe in the control plane.  Figure~\ref{fig:tpp-qsize-example-topo}
shows a six-node dumbell topology on
Mininet~\cite{handigol2012reproducible}, in which each node sends a
small 10kB message to every other node in the topology.  The total
application-level offered load is 30\% of the hosts' network capacity
(100Mb/s).  We instrumented every packet with a TPP, and collected
fully executed TPPs carrying network state at one host.
Figure~\ref{fig:tpp-qsize-example-queues} shows the queue evolution of
6 queues inside the network obtained from every packet received at
that host.

\smallsec{Overheads}:  The actual TPP consists of three instructions,
one each to read the switch ID, the port number, and the queue size,
each a 16 bit integer.  If the diameter of the network is
5~hops, then each TPP adds only a 54 byte overhead to each
packet: 12 bytes for the TPP header~(see \S\ref{subsec:tppformat}),
12~bytes for instructions, and $6\times{}5$~bytes to collect statistics
at each hop.

\subsection{Rate-based Congestion Control}\label{subsec:rcp}
While the previous example shows how TPPs can help observe latency
spikes, we now show how such visibility can be used to {\em control}
network congestion.  Congestion control is arguably a dataplane task,
and the literature has a number of ideas on designing better
algorithms, many of which require switch support.  However, TCP and
its variants still remain the dominant congestion control algorithms.
Many congestion control algorithms, such as
XCP~\cite{katabi2002congestion}, FCP~\cite{han2013fcp},
RCP~\cite{dukkipati2006flow}, etc.\ work by monitoring state that
indicates congestion and adjusting flow rates every few RTTs.

We now show how end-hosts can use TPPs to deploy a new congestion control
algorithm that enjoys many benefits of in-network algorithms, such as Rate Control Protocol
(RCP\@)~\cite{dukkipati2006flow}.  RCP is a congestion control algorithm
that rapidly allocates link capacity to help flows finish quickly.  An
RCP router maintains one fair-share rate $R(t)$ per link (of capacity
$C$, regardless of the number of flows), computed periodically (every
$T$ seconds) as follows:

\begin{equation}\label{eq:rcp}
R(t+T) = R(t)\left(1 - \frac{T}{d}\times\frac{a\;(y(t) - C) + b\;\frac{q(t)}{d}}{C}\right)
\end{equation}

Here, $y(t)$ is the average ingress link utilization, $q(t)$ is the
average queue size, $d$ is the average round-trip time of flows
traversing the link, and $a$ and $b$ are configurable parameters.
Each router checks if its estimate of $R(t)$ is smaller than the
flow's fair-share (indicated on each packet's header); if so, it
replaces the flow's fair share header value with $R(t)$.

We now describe RCP*, an end-host implementation of RCP\@.  The
implementation consists of a rate limiter and a rate controller at
end-hosts for every flow (since RCP operates at a per-flow
granularity).  The network control plane allocates two memory
addresses per link ({\tt Link:AppSpecific\_0} and {\tt
Link:AppSpecific\_1}) to store fair rates.  Each flow's rate
controller periodically (using the flow's packets, or using additional
probe packets) queries and modifies network state in three phases.

\smallsec{Phase 1: Collect.}  Using the following TPP, the rate
controller queries the network for the switch ID on each hop, queue
sizes, link utilization, and the link's fair share rate (and its
version number), for all links along the path.  The receiver simply
echos a fully executed TPP back to the sender.  \cam{The network
updates link utilization counters every millisecond.  If needed,
end-hosts can measure them faster by querying for {\tt
[Link:RX-Bytes]}.}

\begin{verbatim}
PUSH [Switch:SwitchID]
PUSH [Link:QueueSize]
PUSH [Link:RX-Utilization]
PUSH [Link:AppSpecific_0] # Version number
PUSH [Link:AppSpecific_1] # Rfair
\end{verbatim}
\smallsec{Phase 2: Compute.}  In the second phase, each sender
computes a fair share rate $R_{\rm link}$ for each link: Using the
samples collected in phase 1, the rate controller computes the average
queue sizes on each link along the path.  Then, it computes a per-link
rate $R_{\rm link}$ using the RCP control equation.

\smallsec{Phase 3: Update.}  In the last phase, the
rate-controller of each flow asynchronously sends the following TPP to
update the fair rates on all links.  To ensure correctness due to
concurrent updates, we use the {\tt CSTORE} instruction:

\begin{verbatim}
  CSTORE [Link:AppSpecific_0], \
            [Packet:Hop[0]], [Packet:Hop[1]]
  STORE [Link:AppSpecific_1], [Packet:Hop[2]]
PacketMemory:
  Hop1: V_1, V_1+1, R_new_1, (* 16 bits each*)
  Hop2: V_2, V_2+1, R_new_2, ...
\end{verbatim}

\noindent where $V_i$ is the version number in the {\tt AppSpecific\_0} that
the end-host used to derive an updated $R_{{\rm new},i}$ for hop $i$,
thus ensuring consistency.  ({\tt CSTORE dst,old,new} updates {\tt
dst} with {\tt new} only if {\tt dst} was {\tt old}, ignoring the rest
of the TPP otherwise.)  Note that in the TPP, the version numbers
and fair rates are read from packet memory at every hop.


\begin{figure}[t]
\centering
\includegraphics[width=0.5\textwidth]{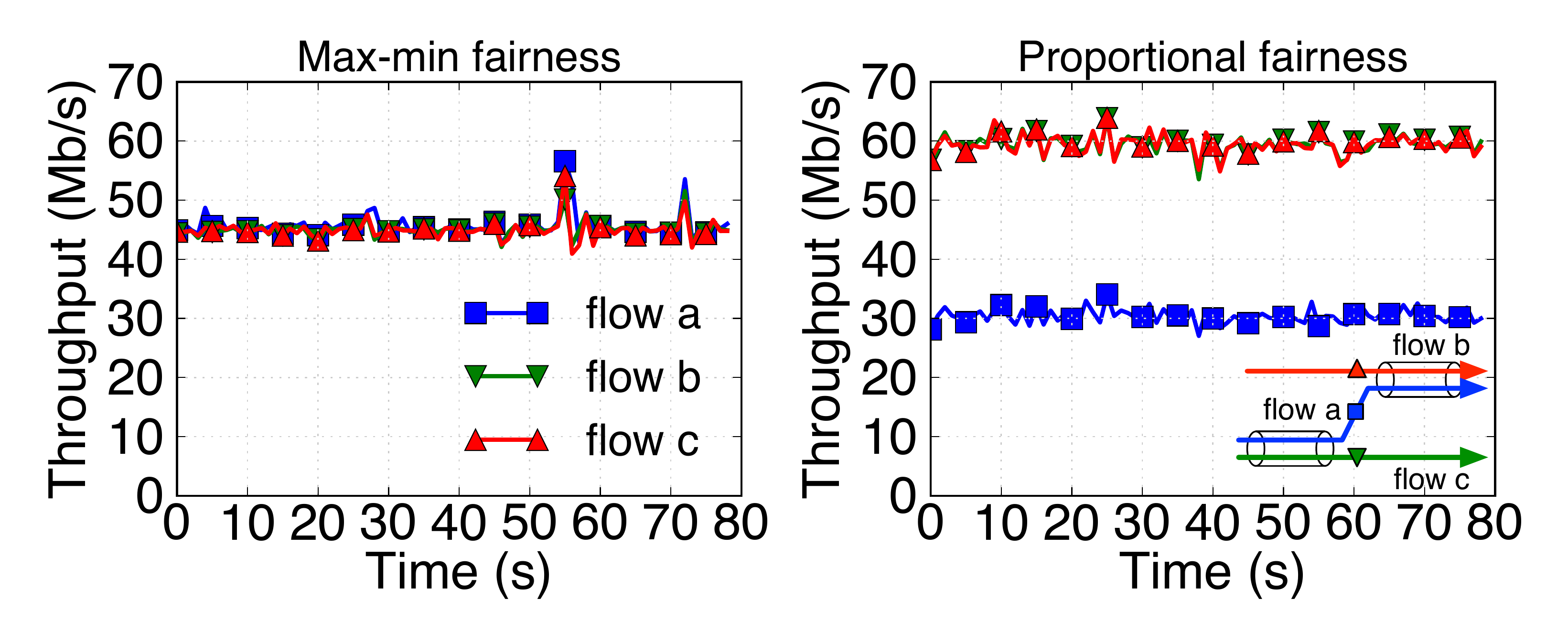}
\caption{Allocations by max-min and proportional fairness variant of RCP
on the traffic pattern shown inset on the right plot; each link has
100Mb/s capacity and all flows start at 1Mb/s at time~0.}
\label{fig:rcp-new}\vspace{-1.2em}
\end{figure}

\smallsec{Other allocations}: Although RCP was originally designed to
allocate bandwidth in a max-min fair manner among competing flows,
Kelly et al.~\cite{kelly2008stability} showed how to tweak RCP to
allocate bandwidth for a spectrum of fairness
criteria---$\alpha$-fairness parameterized by a real number
$\alpha \geq 0$.  $\alpha$-fairness is achieved as follows: if $R_{i}$
is the fair rate computed at the $i$-th link traversed by the flow (as
per the RCP control equation~\ref{eq:rcp}), the flow sets its rate as
\begin{equation}\label{eq:aggregate}
R = \left(\sum\limits_i R_i^{-\alpha}\right)^{-1/\alpha}
\end{equation}

The value $\alpha=1$ corresponds to proportional fairness, and we can
see that in the limit as $\alpha\rightarrow\infty$, $R = \min_i R_i$,
which is consistent with the notion of max-min fairness.  Observe that
if the ASIC hardware had been designed for max-min version of RCP, it
would have been difficult for end-hosts to achieve other useful
notions of fairness.  However, TPPs help defer the choice of fairness
to deployment time, as the end-hosts can aggregate the per-link $R_i$
according to equation~\ref{eq:aggregate} based on \emph{one} chosen
$\alpha$. (We do not recommend flows with different $\alpha$
sharing the same links due to reasons in~\cite{tang2007equilibrium}.)

Figure~\ref{fig:rcp-new} shows the throughput of three flows for both
max-min~RCP* and proportional-fair~RCP* in Mininet: Flow~`a' shares
one link each with flows `b' and `c' (shown inset in the right plot).
Flows are basically rate-limited UDP streams, where rates are
determined using the control algorithm: Max-min fairness should
allocate rates equally across flows, whereas proportional fairness
should allocate
\sfrac{1}{3} of the link to the flow that traverses two links, and
\sfrac{2}{3} to the flows that traverse only one link.

\cam{\smallsec{Overheads}:  For the experiment in Figure~\ref{fig:rcp-new}, the
bandwidth overhead imposed by TPP control packets was about 1.0--6.0\% of
the flows' rate as we varied the number of long lived flows from
3 to 30 to 99 (averaged over 3 runs).  In the same experiment,
TCP had slightly lower overheads: 0.8--2.4\%.  The RCP* overhead
is in the same range as TCP because each flow sends control packets roughly
once every RTT.
As the number of flows $n$ increases, the average per-flow rate decreases
as $1/n$, which causes the RTT of each flow to increase (as the RTT is inversely
proportional to flow rate).  Therefore, the total overhead does not blow up.

\smallsec{Are writes absolutely necessary?}  RCP* is one of the few TPP
applications that \emph{writes} to network state.  It is
worth asking if this is absolutely necessary.  We believe it is necessary
for fast convergence since RCP relies on flows traversing a single bottleneck
link agreeing on \emph{one} shared rate, which is explicitly enforced in RCP.
Alternatively, if rapid convergence
isn't critical, flows can converge to their fair rates in an AIMD fashion
\emph{without} writing to network state.  In fact, XCP implements this AIMD approach, but
experiments in~\cite{dukkipati2006flow} show that XCP converges more
slowly than RCP.}

\subsection{Network Troubleshooting Framework}\label{subsec:netsight}
There has been recent interest in designing programmatic tools for
troubleshooting networks; without doubt, dataplane visibility is
central to a troubleshooter.  For example, consider the task of
verifying that network forwarding rules match the intent specified by
the administrator~\cite{kazemian2013real,khurshid2012veriflow}.  This
task is hard as forwarding rules change constantly, and a
network-wide `consistent' update is not a trivial
task~\cite{reitblatt2012abstractions}.  Verification is further complicated by
the fact that there can be a mismatch between the control plane's view
of routing state and the actual forwarding state in hardware (and such
problems have shown up in a cloud provider's production
network~\cite{chakim-personal}).  Thus, verifying whether packets have
been correctly forwarded requires help from the dataplane.

\begin{figure}[t]
\centering
\includegraphics[width=0.40\textwidth]{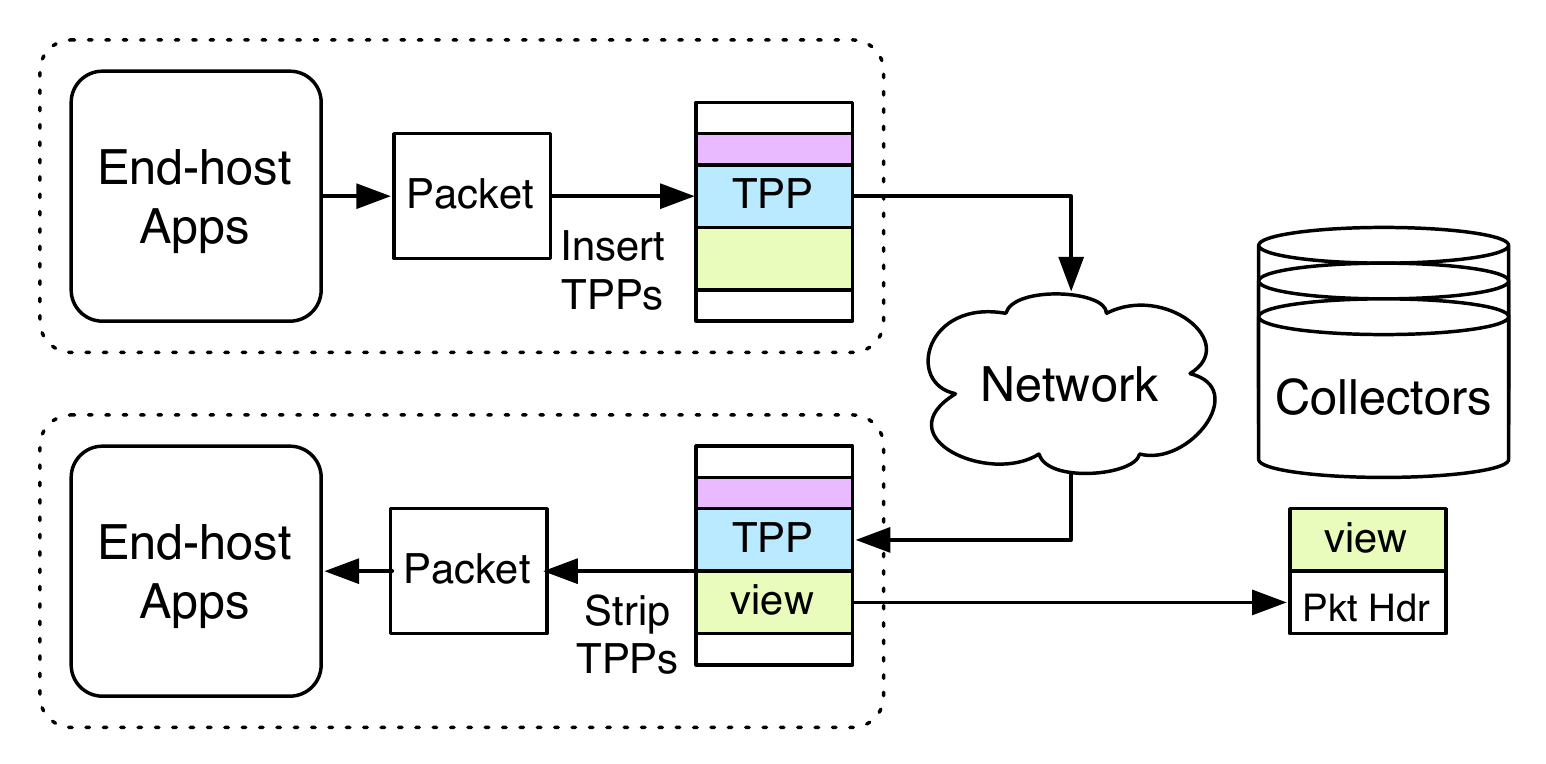}
\caption{TPPs enable end-hosts to efficiently collect
  packet histories, which can then be used to implement four
  different troubleshooting applications described in~\cite{netsight}.}\vspace{-1em}
\label{fig:netsight}
\end{figure}

Recently, researchers have proposed a platform called
NetSight~\cite{netsight}.  NetSight introduced the notion of a `packet
history,' which is a record of the packet's path through the
network and the switch forwarding state applied to the packet.  Using
this construct, the authors show how to build four different network
troubleshooting applications.

We first show how to efficiently capture packet histories that are
central to the NetSight platform.  NetSight works by interposing on
the control channel between the controller and the network, stamping
each flow entry with a unique version number, and modifying flow
entries to create truncated copies of packet headers tagged with the
version number (without affecting a packet's normal forwarding) and
additional metadata (e.g., the packet's input/output ports).  These
truncated packet copies are reassembled by servers to reconstruct the
packet history.

We can refactor the task of collecting packet histories by having a
trusted agent at every end-host (\S\ref{sec:endhost}) insert the TPP
shown below on all (or a subset of) its packets.  On receiving a TPP
that has finished executing on all hops, the end-host gets an accurate
view of the network forwarding state that affected the packet's
forwarding, without requiring the network to create additional packet
copies.
\begin{verbatim}
PUSH [Switch:ID]
PUSH [PacketMetadata:MatchedEntryID]
PUSH [PacketMetadata:InputPort]
\end{verbatim}

Once the end-host constructs a packet history, it is forwarded to
collectors where they can be used in many ways.  For instance, if the
end-host stores the histories, we get the same functionality as {\tt
netshark}---a network-wide {\tt tcpdump} distributed across servers.  From
the stored traces, an administrator can use any query language
(e.g., SQL) to extract relevant packet histories, which gives the same
functionality as the interactive network debugger {\tt ndb}.  Another
application, {\tt netwatch} simply uses the packet histories to verify
whether network forwarding trace conforms to a policy specified by the
control plane (e.g., isolation between tenants).

\smallsec{Overheads}: The instruction overhead is 12 bytes/packet and
6 bytes of per-hop data.  With a
TPP header and space for 10 hops, this is 84 bytes/packet.  If
the average packet size is 1000 bytes, this is a 8.4\% bandwidth
overhead if we insert the TPP on every packet.  If we enable it only
for a subset of packets, the overhead will be correspondingly lower.

\smallsec{Caveats}: Despite its benefits, there are drawbacks to using only TPPs,
especially if the network transforms packets in erroneous or non-invertible ways.
We can overcome dropped packets by sending packets that will be
dropped to a collector (we describe how in~\S\ref{subsec:otherposs}).
Some of these assumptions (trusting the dataplane to function
correctly) are also made by NetSight, and we
believe the advantages of TPPs outweigh its drawbacks.  For instance,
TPPs can collect more statistics, such as link utilization and queue
occupancy, along with a packet's forwarding history.

\begin{ecam}
\subsection{Distributed Load Balancing}\label{subsec:conga}
We now show how end-hosts can use
TPPs to probe for network congestion, and use this detailed visibility to load
balance traffic in a distributed fashion.  We demonstrate a simplified
version of CONGA~\cite{conga}, which is an in-network scheme for
traffic load balancing.  CONGA strives to
maximize network throughput and minimize the maximum network link
utilization in a distributed fashion
by having network switches maintain a table of path-level congestion
metrics (e.g., quantized link utilization).  Using this information,
switches route small bursts of flows (``flowlets'') selfishly
on the least loaded path.  CONGA is optimized for datacenter
network topologies; we refer the curious reader to~\cite{conga}
for more details.
\begin{figure}[t]
\centering
\includegraphics[width=0.45\textwidth]{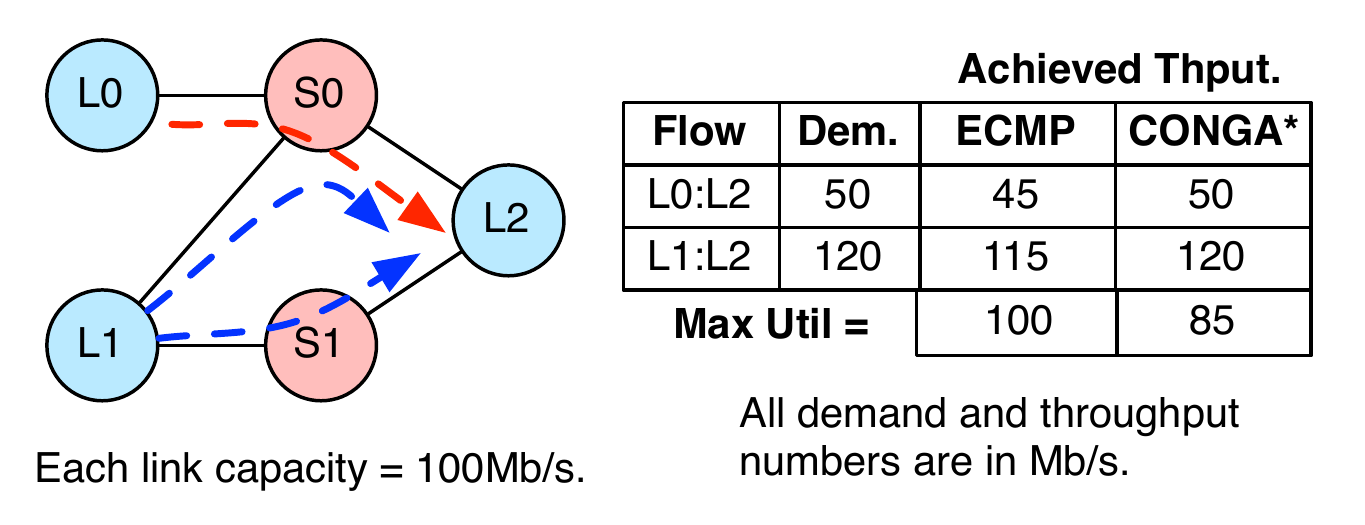}
\caption{An example showing the benefits of congestion-aware load balancing:
ECMP splits flow from L1 to L2
equally across the two paths resulting in suboptimal network utilization.
CONGA*, an end-host refactoring of CONGA~\cite{conga} is able to detect
and reroute flows, achieving optimum in this example.}
\label{fig:conga}\vspace{-1em}
\end{figure}

CONGA's design highlights two benefits relevant to our discussion.
First, it uses explicit visibility by having switches stamp quantized
congestion information on packet headers.  Second, load balancing decisions
are made at round-trip timescales to rapidly detect and react to network
congestion.  Since TPPs also offer similar benefits, we show how we
can refactor the load balancing task between end-hosts and the network,
without requiring custom hardware (except, of course, to support TPPs).

First, we require the network to install multipath routes that
end-hosts can select based on packet header values.  This can be done in
the slow-path by the control plane by programming a `group table' available
in many switches today for multipath routing~\cite[\S5.6.1]{of1.4}, which
selects an output port by hashing on header fields (e.g., the VLAN tag).
This allows end-hosts to select network paths simply by changing the VLAN ID.

Second, we need end-hosts to query for link utilization across various paths,
by inserting the following TPP on a subset of packets destined to hosts
within the datacenter:
\begin{verbatim}
PUSH [Link:ID]
PUSH [Link:TX-Utilization]
PUSH [Link:TX-Bytes]
\end{verbatim}

We query for {\tt Link:TX-Bytes} to measure small
congestion events if the link utilization isn't updated.  The receiver echoes
fully executed TPPs back to the
sender to communicate the congestion.  Note that the header of the echoed TPP
also contains the path ID along with the link utilization on each link in the path.

Third, using information in the fully executed TPPs, end-hosts can build a
table mapping `Path~$i\rightarrow$ Congestion~Metric ($m_i$),' where $m_i$
is either the maximum or sum of link utilization on each switch--switch
network hop on path $i$.  The authors of CONGA note that `sum' is closer to
optimal than `max' in the worst-case scenario (adversarial);
however CONGA used `max' as it does not cause overflows when switches aggregate
path-congestion.  With TPPs, this is not an issue, and the choice can be deferred
to deploy time.

And finally, end-hosts have full context about flows and flowlets, and
therefore each end-host can select a flowlet's path by setting the
path tag appropriately on the flowlet's packets.

\smallsec{Overheads}:  We implemented a proof-of-concept prototype (CONGA*) in software using UDP flows;
Figure~\ref{fig:conga} reproduces an example from CONGA~\cite[Figure~4]{conga}.
We configured switches S0 and S1 to select paths based on destination UDP
port.  The flow from L0 to L2 uses only one path, whereas the
flow from L1 to L2 has two paths.  The UDP agents at L0 and L1 query
for link utilization and aggregate congestion metrics every millisecond
for the two paths.  With CONGA*, end-hosts can maximize network
throughput meeting the demands for both flows, while simultaneously minimizing
the maximum link utilization.  In this example, the overhead introduced by
TPP packets was minimal ($< 1\%$ of the total traffic).

\smallsec{Remark}: Note that the functionality is \emph{refactored}
between the network and end-hosts; not all functionality resides completely
 at the end-hosts.  The network implements TPP and multipath
routing.  The end-hosts merely select paths based on congestion completely
in software.
\end{ecam}

\begin{eext}
\subsection{Low-overhead Measurement}\label{subsec:sketch}
Since TPPs can read network state, it is straightforward to use them
to monitor the network.  However, we show how to implement non-trivial
measurement tasks.  In particular, OpenSketch~\cite{yu2013software} is
a recently published measurement framework that makes the
observation that many measurement tasks can be approximated accurately
using probabilistic summary algorithms called ``sketches,'' which can in turn be compiled down to a three-stage
pipeline where packets are hashed, filtered, and counted.  The authors
show how to combine these primitives to answer questions such as: (i)
what is the number of unique IP addresses communicating across links?
(ii) what is the flow size distribution across switches?

An example of a sketch is the bitmap sketch, which can count the
number of unique elements in a stream as follows: hash the element
(e.g.\ source IP address) to one of $b$ bits and set it to 1.  The
estimate of the cardinality of the set is $b\times\ln(b/z)$, where $z$
is the number of unset bits~\cite{estan2003bitmap}.

Many sketches require multiple hash functions in hardware operating at
line-rate.  This introduces a new ASIC functionality that is specific
to sketches; thus, it is worth asking if the task can be refactored
using the visibility offered by TPPs.  Since end-hosts can readily
implement many hash functions cheaply in software, the only piece of
information they are missing is the packet's routing context, which
can be obtained by the following TPP:
\begin{verbatim}
PUSH [Switch:ID]
PUSH [PacketMetadata:OutputPort]
\end{verbatim}

As example, consider a task where one wants to measure the number of
unique destination IP addresses traversing the core switches in the
network, and update these values every few (say 10) seconds.  Each
end-host in the network participates in this task by inserting the
above TPP into its packets.  Note that this TPP need not be inserted
into all packets, but it should be inserted at least once for every
destination IP address the host communicates with.  The receiving
end-host parses the fully executed TPP to retrieve the (switch,link)
IDs from the packet and implements the sketch as follows:

\begin{verbatim}
index = hash(packet.ip.dest)
foreach (switch,link) in tpp:
  bitmask[switch][index] = 1
\end{verbatim}

\begin{figure}[t]
\centering
\includegraphics[width=0.45\textwidth]{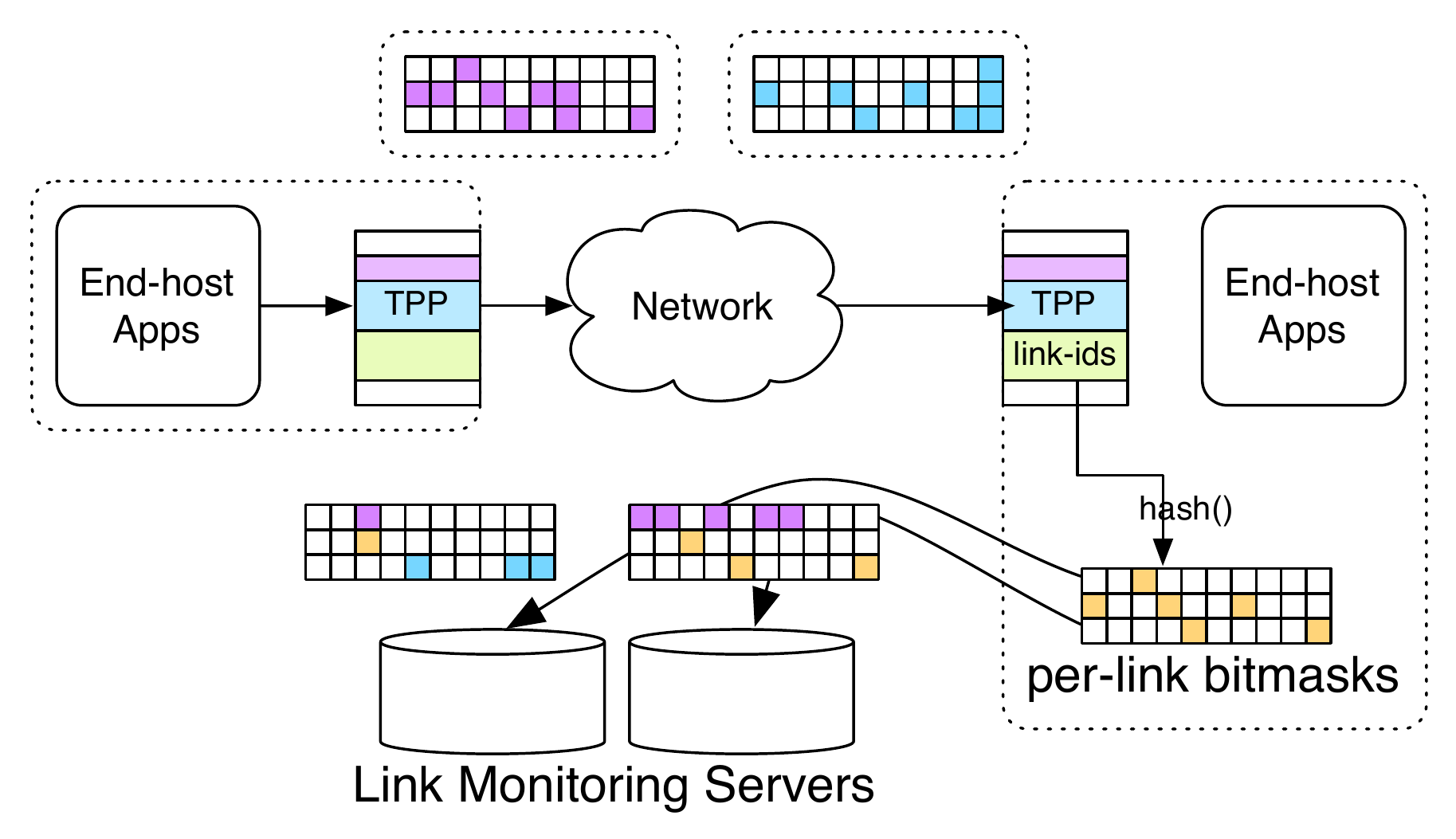}
\caption{Refactoring the bitmap sketch~\cite{estan2003bitmap} to estimate set cardinality.}
\label{fig:opensketch}
\end{figure}

The sketch data-structures are distributed across the end-hosts in the
network, but we can take advantage of the fact that the sketch
operation (`bit-set') is commutative.  Every 10~seconds, the end-hosts push
those summary data structures that have changed since the last
interval to a central, load-balanced, link monitoring service.  The
link monitoring service aggregates the bit-vectors to obtain the
sketch data-structure for every link, obtaining the same result as one
would obtain using OpenSketch.  This refactoring allow end-hosts to
retain flexibility in implementing other kinds of sketches.

\smallsec{Overheads}: To implement the count-cardinality sketch, we
only need one TPP per unique destination IP address.  If we sample one
out of every 10 packets to insert the measurement TPPs, we will incur
less than 1\% bandwidth overhead due to extra headers in packets.  If
we assume a $k=64$~FatTree datacenter network, there are 65536 core
links, and 65536~servers.  The sketch data-structure's accuracy depends
on the number of bits and the probability of
collision~\cite{estan2003bitmap}.  If we use 1kbit memory per link,
the total memory usage for all 65536 links is about 8MB/server.
\end{eext}

\subsection{Other possibilities}\label{subsec:otherposs}
The above examples illustrate how a single TPP interface
enables end-hosts to achieve many tasks.  \cam{
There are more tasks that we couldn't cover in detail.  In the interest
of space, we refer the reader to the extended version of this paper
for more details on some of the tasks below~\cite{tpp-extended}.}

\cam{\smallsec{Measurement}: Since TPPs can read network state, they
can be used in a straightforward fashion for measuring any network
statistic at rapid timescales.  As TPPs operate in the dataplane,
they are in a unique position to expose path characteristics
experienced by a specific packet that an end-host cares about.}

\ignore{\smallsec{Source routing and load balancing}: Other simple examples include
source-routing, which enables packets to control routes to the
destination.  This is achieved if the TPP overwrites the {\tt
PacketMetadata:OutputPort} register.  Moreover, control over routes
also enables randomized load balancing, where end-hosts can use TPPs
to probe the link utilization on alternate packet routes to the
destination to effectively distribute load across multiple network
links~\cite{mitzenmacher2001power}.  Other load balancing proposals
such as Hedera~\cite{al2010hedera}, Dahu~\cite{radhakrishnan2013dahu},
LocalFlow~\cite{senscalable}, etc.\ share a common
monitor-control-modify workflow, which can be efficiently implemented
with TPPs, without being limited by the control plane's slow response
time.}


\smallsec{Network verification}:  TPPs also help in verifying whether
network devices meet certain requirements.  For example, the path
visibility offered by TPPs help accurately verify that route
convergence times are within an acceptable value.  This task can be
challenging today, if we rely on end-to-end reachability as a way to
measure convergence, because backup paths can still maintain
end-to-end connectivity when routes change.  Also, the explicit
visibility eases fault localization~\cite{zeng2012automatic}.

\smallsec{Fast network updates}:  By allowing secure applications to write
to a switch's forwarding tables, network updates can be made very
fast.  This can reduce the time window of a transient state when
network forwarding state hasn't converged.  For example, it is
possible to add a new route to all switches along a path in half a
round-trip time, as updating an IP forwarding table requires only
64~bits of information per-hop: 32~bit address and a 32~bit netmask
per hop, tiny enough to fit inside a packet.

\smallsec{Wireless Networks}:  TPPs can also be used in wireless
networks where access points can annotate end-host packets with
rapidly changing state such as channel SNR\@.  Low-latency access to
such rapidly changing state is useful for network diagnosis, allows
end-hosts to distinguish between congestive losses and losses due to
poor channel quality, and query the bitrate that an AP selected for a
particular packet.
\section{Design of TPP-Capable Switches}
\label{sec:design}
\begin{figure*}[t]
\centering
\includegraphics[width=0.95\textwidth]{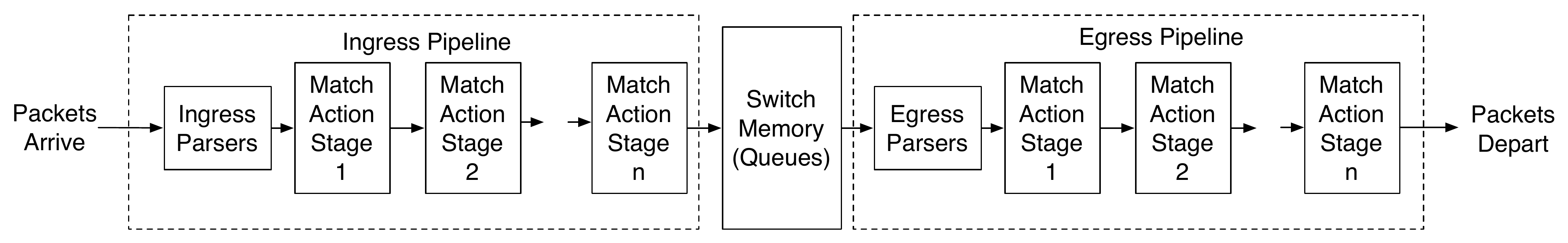}
\caption{A simplified block diagram of the dataplane pipeline in a
  switch ASIC.  Packets arrive
  at the ingress, and pass through multiple modules.  The scheduler
  then forwards packets (that are not dropped) to the output ports
  computed at earlier stages in the pipeline.}
\label{fig:pipeline}
\end{figure*}
\cam{
In this section, we discuss the TPP instructions, addressing schemes,
and the semantics of the TPP interface to
a switch and what it means for a switch to be TPP-capable.  Network
switches have a variety of form factors and implementations; they
could be implemented in software (e.g., Click, Open~vSwitch), or in
network processors (e.g., NetFPGA), or as hardware ASICs.
A switch might also be built hierarchically from multiple
ASICs, as in `chassis' based switches~\cite[Figure~3]{jain2013b4}.
A TPP can be executed on each of these platforms.
Thus, it is useful for a TPP-capable switch and the end-host
to have a contract that preserves useful properties 
without imposing a performance penalty.  We achieve this
by constraining the instruction execution order and atomicity.}


%
\subsection{Background on a Switch Pipeline}
We begin with an abstract
model of a switch execution environment shown in Figure~\ref{fig:pipeline}.
The packet flows from input to output(s) through many pipelined
modules.  \cam{Once a packet arrives at an input port, the
dataplane tags the packet with
metadata (such as its ingress port number)}.  Then, the packet passes
through a parser that extracts fields from the packet
and passes it further down the pipeline which consists of several
match-action stages.  This is also known as multiple match table
model~\cite{glen2013hardware}.  For example, one stage might use the
parsed fields to route the packet (using a combination of layer 2 MAC
table, layer 3 longest-prefix match table, and a flexible TCAM table).
Finally, any modifications to the packet are committed and the packet
is queued in switch memory.  Using metadata (such as the packet's
priority), the scheduler decides when it is time for the packet to be
transmitted out of the egress port determined earlier in the pipeline.
The egress stage also consists of a number of match-action stages.  

\cam{\subsection{TPP Semantics}
The read/write instructions within a TPP access two distinct memory spaces: memory within
the switch (switch memory), and a per-hop scratch space within the packet
(packet memory).  By all switch memory, we only mean memory at the stages traversed by a
TPP, except the memory that stores packet contents.  By all packet memory, we mean the
TPP related fields in the packet.
Now, we state our requirements for read/write instructions accessing the
two memory spaces.

\smallsec{Switch memory}: To expose statistics pertaining to a specific packet
as it traverses the network, it is important for the instructions in the TPP
to have access to the same values that are used to forward the packet.  For
read-only values, this requirement means that reads by a TPP to a single memory location
must necessarily be atomic and after all writes
by the forwarding logic to the \emph{same} memory location.
For example, if a TPP accesses the memory that holds the output port
of a packet, it must return the same port that the forwarding
logic determines, and no other value.  This is what we mean by a
``packet-consistent'' view of network state.

For read-write memory addresses, it is
useful if instructions within the TPP were executed in the order specified
by the TPP to a given location \emph{after} any modifications
by the switch forwarding logic.  Thus, writes by a TPP supersede
those performed by forwarding logic.

\smallsec{Packet memory}: Since instructions can read from and write to
packet memory using {\tt PUSH} and {\tt POP}, writes
to packet memory must take effect sequentially in the
order specified by the TPP.  This guarantees that if a TPP pushes values
at memory locations X, Y, and Z onto packet memory, the end-host sees the values
in the packet in the same order.  This does not require that reads to
X, Y, and Z be issued in the same order.}

\begin{figure*}
\centering\vspace{-1em}
\begin{subfigure}[b]{0.35\textwidth}
\includegraphics[width=\textwidth]{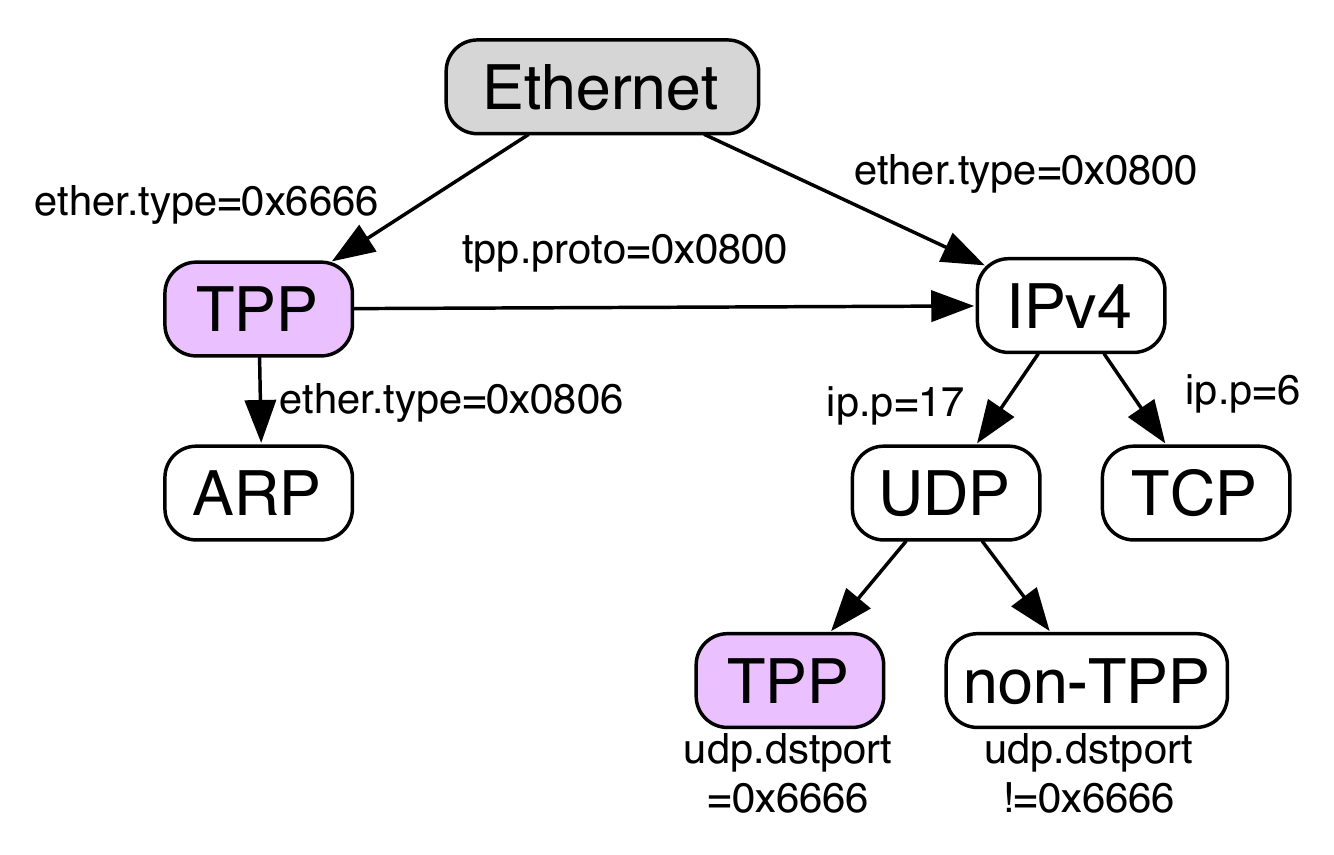}
\caption{Parse graph for the two ways to parse TPPs: transparent mode,
  or standalone mode.}
\label{fig:parsegraph}
\end{subfigure}%
\quad
\begin{subfigure}[b]{0.45\textwidth}
\includegraphics[width=\textwidth]{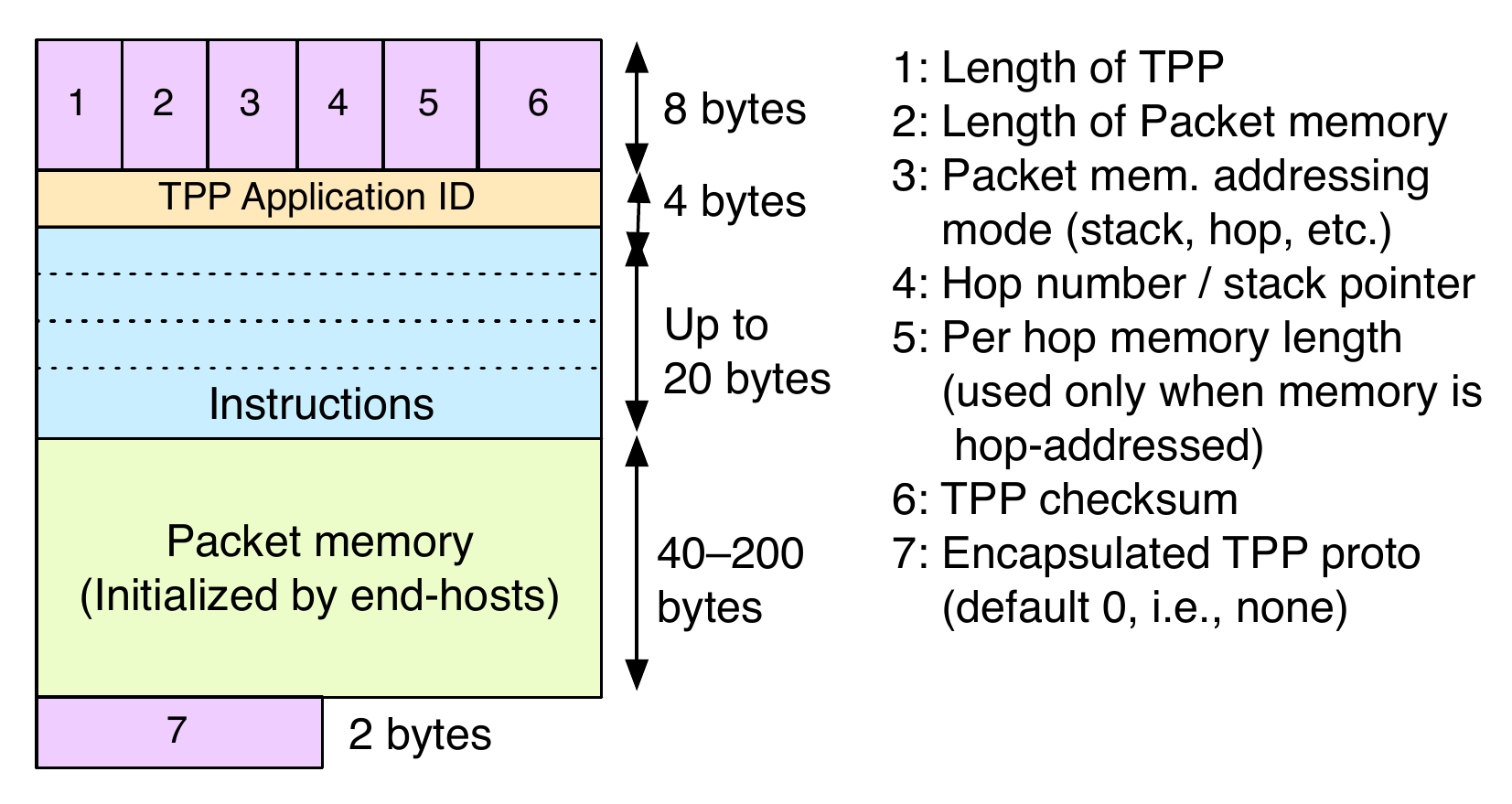}
\caption{TPP's packet structure.}
\label{fig:tpp-packet}
\end{subfigure}
\caption{The parse graph and structure of a TPP\@.  We chose {\tt
    0x6666} as the ethertype and source UDP port that uniquely
  identifies a TPP\@.  With a programmable switch parser, this choice
  can be reprogrammed at any time.}\vspace{-1em}
\end{figure*}

\subsection{TPP Execution Model}\label{subsec:model}
TPPs are executed in the dataplane pipeline.
TPPs are required to fit exactly within an MTU to avoid having the
ASIC deal with fragmentation issues.  This is not a big limitation, as
end-hosts can split a complex task into multiple smaller TPPs if a
single packet has insufficient memory to query all the required
statistics.  By default, a TPP executes at every hop, and instructions
are not executed if they access memory that doesn't exist. \cam{This
ensures the TPP fails gracefully.

Furthermore, the platform is free to reorder reads and writes
so they execute in any order.  However, if the programmer
needs guarantees on ordering instructions due to data hazards (e.g., for
{\tt CEXEC, CSTORE}), they must ensure the TPP accesses memory in the
pipeline order.  For a vast majority of use cases, we argue this
restriction is not severe.  From a practical standpoint, this requirement
ensures that the switch pipeline remains feed-forward as it is today in
a majority of switches.}

\subsubsection{Unified Memory-Mapped IO}\label{subsubsec:mmio}
A TPP has access to any statistic computed by the switch that is addressable.  The
statistics can be broadly namespaced into per-switch (i.e., global),
per-port, per-queue and per-packet.  Table~\ref{tab:stats} shows
example statistics in each of these namespaces.  These statistics may
be scattered across different stages in the pipeline, but TPPs access
them via a unified address space.  For instance,
a switch keeps metadata such as input port, the selected route, etc.\ for
every packet that can be made addressable.  These address
mappings are known upfront to the TPP compiler that converts mnemonics
such as {\tt [PacketMetadata:InputPort]} into virtual addresses.

\subsubsection{Addressing Packet Memory}
Memory is managed using a stack pointer and a {\tt PUSH}
instruction that appends values to preallocated packet memory.  TPPs
also support a hop addressing scheme, similar to the the {\tt
  base:offset} x86-addressing mode.
Here, {\tt base:offset} refers
to the word at location {\tt base * hop\_size + offset}.  Thus, if hop-size is
16 bytes, the instruction ``{\tt LOAD [Switch:SwitchID],
  [Packet:hop[1]]}'' will copy the switch ID into {\tt PacketMemory[1]}
on the first hop, {\tt PacketMemory[17]} on the second hop, etc.\ The
{\tt offset} is part of the instruction; the {\tt base} value (hop number)
and per-hop memory size values are in the
TPP header.  To simplify memory management in the dataplane, the end-host
must preallocate enough space in the TPP to hold per-hop data structures.

\subsubsection{Synchronization Instructions}
Besides read and write, a useful instruction in a concurrent programming
environment is an atomic update instruction, such as a conditional
store {\tt CSTORE}, conditioned on a memory location matching a
specified value, \cam{halting subsequent instructions in the TPP if the
update fails.  That is, {\tt CSTORE [X],[Packet:hop[Pre]],[Packet:hop[Post]]}
works as follows:}
\begin{verbatim}
  succeeded = False
  if (value at X == value at Packet:hop[Pre]) {
    value at X = value at Packet:hop[Post]
    succeeded = True
  }
  value at Packet:hop[Pre] = value at X;
  if (succeeded) {
    allow subsequent instructions to execute
  }
\end{verbatim}

\cam{By having {\tt CSTORE} return the value of {\tt X}, an end-host can
infer if the instruction succeeded.  Notice that the second and third
operands are read from a unique location at every hop.
This is needed to ensure correct semantics when the switch overwrites
the value at the second operand.}

In a similar vein, we found a conditional execute ({\tt CEXEC}) instruction useful;
for example, it may be desirable to execute a
network task only on one switch, or on a subset of switches (say all
the top of rack switches in a datacenter).  The conditional execute
instruction specifies a memory address,
a 32-bit mask, and a 32-bit value (specified in
the packet hop), which instructs the switch to execute all subsequent
instructions only when {\tt (switch\_value \& mask) == value}.  All
instructions that follow a failed {\tt CEXEC} check will not be
executed.
\subsection{Parsing: TPP Packet Format}\label{subsec:tppformat}
As noted in \S\ref{sec:examples}, a TPP is any Ethernet frame from
which we can uniquely identify a TPP header, the instructions, packet
memory, and an optional payload.  This allows end-hosts to use TPPs in
two ways: (i) piggy-back TPPs on any existing packet by encapsulating
the packet within a TPP of ethertype {\tt 0x6666}, or (ii) embed a TPP
into an otherwise normal UDP packet destined for port {\tt 0x6666},
which is a special port number usurped by TPP-enabled routers.

Figure~\ref{fig:parsegraph} shows the two parse graphs depicting the
two ways in which our prototype uses TPPs.  A parse graph depicts a
state machine for a packet parser, in which the nodes denote protocols
and edges denote state transitions when field values match.  We use
the same convention as in~\cite{glen2013hardware} to show the two ways
in which we can parse TPPs.

\subsection{Putting it together: the TCPU}\label{subsec:tcpu}
TPPs execute on a tiny processor, which we call the TCPU.
A simple way to implement the TCPU is by having a RISC-like processor
at the end of the ingress match-action
stages \cam{as we described in our earlier position paper~\cite[Figure~5]{jeyakumar2013tiny}}.
\cam{This simple approach could be practical for software, or low-speed
hardware switches, but might be impractical in high-speed hardware switches
as memory in an ASIC is often distributed across modules.}  The wiring complexity
to provide read and write paths from each module to the TCPU becomes prohibitively
expensive within an ASIC, and is simply infeasible across line-cards in a chassis
switch.

\begin{figure}[t]
\centering
\includegraphics[width=0.5\textwidth]{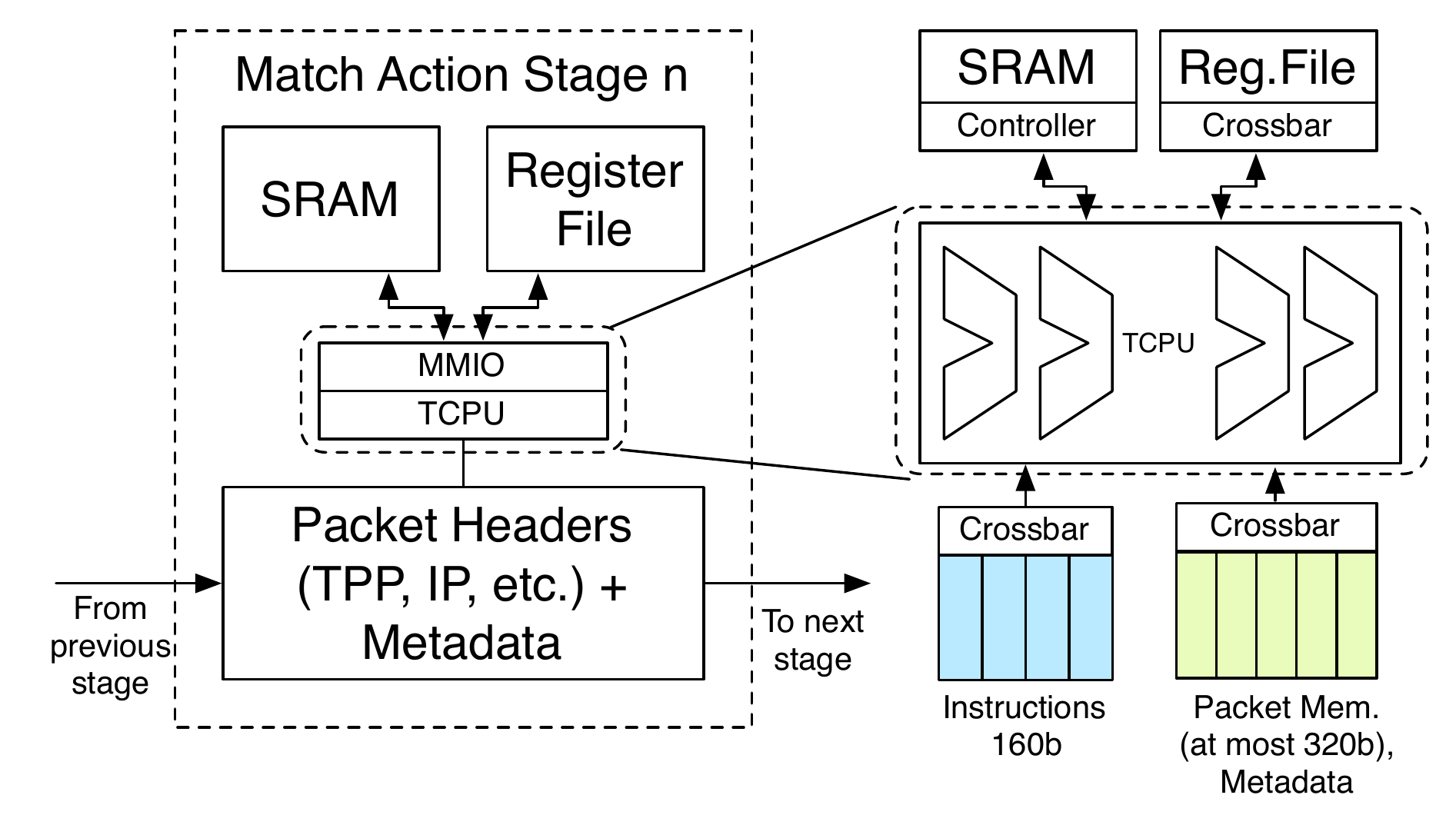}
\caption{At every stage, the TCPU has execution units that can access
  only local memory and registers, as well as packet metadata.}
\label{fig:tcpu}
\end{figure}

\cam{We overcome this limitation in two ways.  First, our execution
model permits reordering reads and writes across different ASIC memory
locations.  Second, end-hosts can statically analyze a desired TPP and split it
into smaller TPPs if one TPP is insufficient.  For instance, if an end-host
requires link utilization on all links at all switches a packet traverses,
it can stage the following sequence of TPPs: (i) send one TPP to collect
switch ID and link utilizations on links traversed by the packet, and (ii)
send a new TPP to each switch link on the switches traversed by TPP~1
to collect the remaining statistics.  To summarize:}

\begin{itemize}[noitemsep,leftmargin=1em]
\item Loads and stores in a single packet can be executed in any
  order, by having end-hosts ensure there are no write-after-write, or
  read-after-write conflicts.
\item The operands for conditional instructions, such as {\tt CSTORE}
  and {\tt CEXEC}, are available before, or at the stages where the
  subsequent instructions execute; {\tt CEXEC} can execute when all
  its operands are available.
\end{itemize}

By allowing instructions to be executed out of order, we can
distribute the single logical TCPU \cam{on an ASIC} by replicating its
functionality at every stage.
Each stage has one execution unit for every instruction in the packet,
a crossbar to connect the execution units to all registers local to
the stage and packet memory, and access to the stage's local memory
read/write port.  From the decoded instructions, the stage can execute
all instructions local to the stage, and once all memory accesses have
completed, the packet leaves the stage.

Replicating execution units might seem expensive, but the majority of
logic area in an ASIC is due to the large memories (for packet
buffers, counters, etc.), so the cost of execution units is not
prohibitive~\cite{glen2013hardware}.  Figure~\ref{fig:tcpu} shows the
TCPU if we zoom into one of the match-action stages.

\cam{\smallsec{Serializing PUSH/POP instructions}: Finally, there are
many techniques to ensure the effect
of {\tt PUSH} and {\tt
  POP} instructions appear if they executed inorder.}  Since the packet memory
  addresses accessed by {\tt PUSH/POP}
instructions are known immediately when they are parsed, they can be
converted to equivalent {\tt LOAD}/{\tt STORE}s that can then be executed
out of order.  For example, consider the following TPP:\vspace{-0.5em}

{\fontsize{9pt}{3pt}
\begin{verbatim}
PUSH [PacketMetadata:OutputPort]
PUSH [PacketMetadata:InputPort]
PUSH [Stage1:Reg1]
POP  [Stage3:Reg3]
\end{verbatim}
}

After parsing the instructions, they can be converted to the following
TPP which is equivalent to the above TPP:

{\fontsize{9.2pt}{3pt}
\begin{verbatim}
LOAD  [PacketMetadata:OutputPort], [Packet:Hop[0]]
LOAD  [PacketMetadata:InputPort], [Packet:Hop[1]]
LOAD  [Stage1:Reg1], [Packet:Hop[2]]
STORE [Stage3:Reg3], [Packet:Hop[2]]
\end{verbatim}
}
Now, the TPP loads the values stored in two registers to the packet memory
addressed in the hop addressing format.  Note that the packet's output
port is not known until the packet is routed, i.e., at the end of the
ingress stage.  The execution proceeds as follows:
\begin{itemize}[noitemsep,leftmargin=1em]
  \item By ingress stage 1, the metadata consists of four instructions,
    the memory addresses they access (the four registers and the three
    packet memory offsets), the packet's hop number, the packet's
    headers, its input port, its CRC, etc.
  \item At stage 1, the packet's input port is known.  Stage 1
    executes the second instruction, and stores the input port value
    at the 2nd word of the packet memory.  Stage 1 also executes the
    third instruction, copying {\tt Reg1} to the 3rd word of packet
    memory.
  \item At stage 3, the fourth instruction executes, copying the 3rd
    word from packet memory into {\tt Reg3}.
  \item At the end of the ingress stage, the packet's output port is
    already computed, and the last stage copies the output port number
    to the 1st word of the packet memory before the packet is stored
    in the ASIC packet buffers.
\end{itemize}

\section{End-host Stack}
\label{sec:endhost}
Now that we have seen how to design a TPP-enabled ASIC, we look at the
support needed from end-hosts that use TPPs to achieve a complex
network functionality.  Since TPP enables a wide range of applications
that can be deployed in the network stack (e.g., RCP congestion
control), or individual servers (e.g., network monitoring), or a
combination of both, we focus our efforts on the
common usage patterns.
\begin{figure}[t]
\centering
\includegraphics[width=0.45\textwidth]{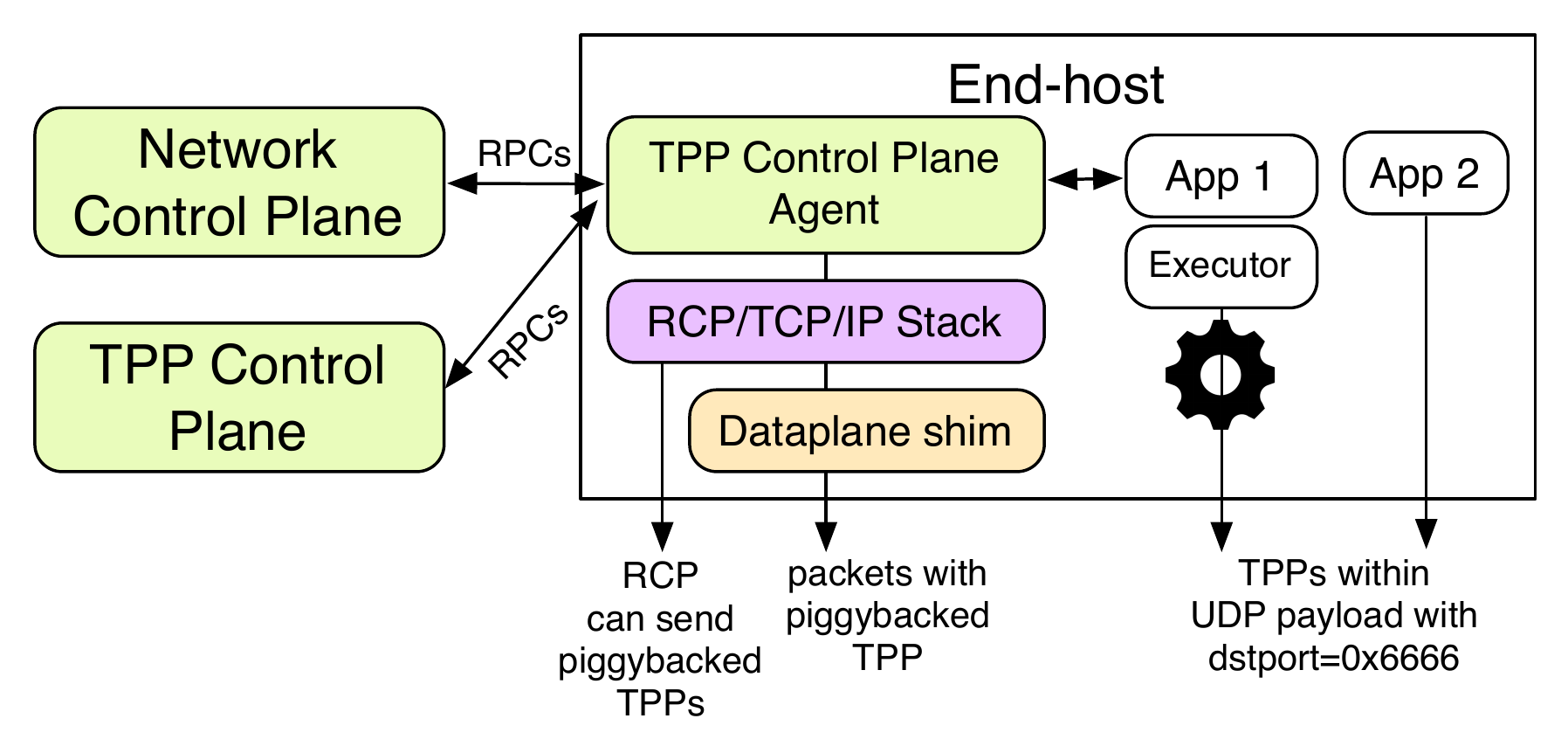}
\caption{End-host stack for creating and managing TPP-enabled
  applications.  Arrows denote packet flow paths through the
  stack, and communication paths between the end-host and the
  network control plane.}\vspace{-1em}
\label{fig:endhost}
\end{figure}

\smallsec{End-host architecture}: The purpose of the end-host stack (Figure~\ref{fig:endhost}) is
to abstract out the common usage patterns of TPPs and implement TPP
access control policies.  At every end-host, we have a TPP control-
and dataplane agent.  The control plane is a software agent that does
not sit in the critical forwarding path, and interacts with the
network control plane if needed.  The dataplane shim sits on the
critical path between the OS network stack and the network interface
and has access to every packet transmitted and received by the
end-host.  This shim is responsible for transparently adding and
removing TPPs from application-generated packets, and enforcing access
control.

\subsection{Control plane}
The TPP control plane (TPP-CP) is a central entity to keep track of
running TPP applications and manage switch memory, and has an agent at
every end-host that keeps track of the active TPP-enabled applications
running locally.  Each application is allocated a contiguous set of
memory addresses that it can read/write.  For example, the RCP
application requires access to a switch memory word to store the
$R_{\rm fair}$ at each link, and it owns this memory exclusively.
This memory access control information is similar to the x86's global
descriptor table, where each entry corresponds to a segment start and
end address, and permissions to read/write to memory is granted
accordingly.

TPP-CP exports an API which authorized applications can use to insert
TPPs on a subset of packets matching certain criteria, with a certain
sampling frequency.  Note that TPP-CP will know the caller application
(e.g. {\tt ndb}) so it can deny the API call if the TPP accesses
memory locations other than those permitted.  The API definition is as
follows:

{\small
\begin{verbatim}
  add_tpp(filter, tpp_bytes, sample_frequency, priority)
\end{verbatim}
}%
\noindent where {\tt filter} is a packet filter (as in {\tt
  iptables}), and {\tt tpp\_bytes} is the compiled TPP, and {\tt
  sample\_frequency} is a non-negative integer that indicates the
sampling frequency: if it is $N$, then a packet is stamped with the
TPP with probability $1/N$.  If $N=1$, all packets have the TPP\@.
The dataplane stores the list of all TPPs with each filter: This
ensures that multiple applications, which want to install TPPs on
(say) 1\% of all IP packets, can coexist.

TPP-CP also configures the dataplane to enforce access control
policies.  Each memory access policy is a tuple: {\tt
  (appid,op,address\_range)}.  The value {\tt appid} is a 64-bit
number, {\tt op} is either {\tt read} or {\tt write}, and {\tt
  address\_range} is an interval denoting the start and end address.
The TPPs are statically analyzed, to see if it accesses memories
outside the permitted address range; if so, the API call returns a
failure and the TPP is never installed.

\subsection{Dataplane}
\cam{The end-host dataplane is a software packet processing pipeline
that allows applications to inject TPPs into ongoing packets, process
executed TPPs from the network, and enforce access control policies.}

\smallsec{Interposition}: The dataplane realizes the TPP-CP API {\tt
  add\_tpp}.  It matches outgoing packets against the table of filters
and adds a TPP to the first match, or sends the packet as such if
there is no match.  Only one TPP is added to any packet.  The
interposition modules in the dataplane also strips incoming packets
that have completed TPPs before passing the packet to the network
stack, so the applications are oblivious to TPPs.


\smallsec{Processing executed TPPs}: The dataplane also processes
incoming packets from the network, which have fully executed.  It
echoes any standalone TPPs that have finished executing back to the
packet's source IP address.  For piggy-backed TPPs, the dataplane checks
the table mapping the application ID to its aggregator, and sends the
finished TPP to the application-specific aggregator.

\subsection{Security considerations}
\label{subsec:security}
There is a great deal of software generating network traffic in any
datacenter, and most of it should not be trusted to generate
arbitrary TPPs.  After all, TPPs can read and write a variety of
switch state and affect packet routing.  This raises the questions of
how to restrict software from generating TPPs, but also how to provide
some of the benefits of TPPs to software that is not completely
trusted.  We now discuss possible mechanisms to enforce such restrictions,
under the assumption that switches are trusted, and there is a trusted
software layer at end hosts such as a hypervisor.

Fortunately, restricting TPPs is relatively simple, because it boils
down to packet filtering, which is already widely deployed.  Just as
untrusted software should not be able to spoof IP addresses or VLAN
IDs, it should not able to originate TPPs.  Enforcing this restriction
is as easy as filtering based on protocol and port numbers, which can
be done either at all ingress switch ports or hypervisors.

In many settings, read-only access to most switch state is harmless.
(A big exception is the contents of other buffered packets, to which
TPPs do not provide access anyway.)  Fortunately, TPPs are relatively
amenable to static analysis, particularly since a TPP contains at most
five instructions.  Hence the
hypervisor could be configured to drop any TPPs with write
instructions (or write instructions to some subset of switch state).
Alternatively, one could imagine the hypervisor implementing higher-level
services (such as network visibility) using TPPs and expose them to
untrusted software through a restricted API.

At a high level, a compromised hypervisor sending malicious TPPs is as
bad as a compromised SDN controller.  The difference is that
hypervisors are typically spread throughout the datacenter on every
machine and may present a larger attack surface than SDN controllers.
Hence, for defense in depth, the control plane needs the ability to
disable write instructions ({\tt STORE}, {\tt CSTORE}) entirely.
A majority of the tasks we presented required only read access to
network state.

\subsection{TPP Executor}\label{subsec:executor}
Although the default way of executing TPP is to execute at all hops from
source to destination, we have built a `TPP Executor' library that abstracts
away common ways in which TPPs can be (i) executed reliably, despite TPPs
being dropped in the network, (ii) targeted at one switch, without incurring
a full round-trip from one end-host to another, (iii) executed in a scatter-gather
fashion across a subset of switches, and many more.  \nextended{In the interest of space,
we defer a detailed discussion to the extended version of this paper~\cite{tpp-extended}.}
\extended{The goal of this section is to show
how complex operations can be built from a simple set of primitives.

\smallsec{Reliable execution}: Since TPPs are forwarded just like
regular packets, they can be dropped if there is congestion.  The
helper function abstracts away retrying (for a maximum number of
times) before giving up.  Applications can use this functionality for
idempotent operations (e.g., loads and conditional stores).  We can
make stores idempotent by first reading the value and conditioning on
the value for subsequent retries.

\smallsec{Targeted execution}: TPPs can be crafted so that they
execute only at one specific switch.  This helper function wraps a TPP
with a {\tt CEXEC} instruction conditioned on the switch ID matching
the specified value.  The end-host agent creates a UDP packet and
sends it to the switch IP (obtained from the network control plane).

As we noted in \S\ref{sec:design}, a switch might have multiple
pipelines connecting a set of interfaces.  Since interfaces typically
have their own IP addresses, the end-host can send a TPP addressed to a specific
interface which would result in the TPP being routed through a
pipeline.  Furthermore, switches can be configured to reflected back to
the source, using the following execution pattern.

\smallsec{Reflective TPP}: Consider an example where a server wants to
monitor congestion at the top-of-rack switch it is connected to.
Sending a TPP from the server to a destination and back incurs a full
round-trip time.  To achieve even quicker response, the server can
program the switch to reflect specially marked TPPs (in its header)
back to to the source address.  With programmable protocol
parsing~\cite[\S4]{bosshart2013programming}, switches can swap the
source and destination IP addresses inside the TPP so the packet never
leaves the rack.

\smallsec{Scatter gather}: Some monitoring applications collect
statistics from a number of switches, so we provide an executor that
implements scatter-gather with retries.  The application specifies the
list of switches to execute a TPP, and the executor library takes care
of creating the necessary TPPs, and masks any failures.  The control
plane can minimize the number of packets sent from the application to
the switches, by constructing a multicast tree between the sender and
monitored switches (in the slow path).


\smallsec{Large TPPs}: If the statistics collected do not fit into
a packet, either because the number of hops is large, or the number of
statistics collected per-hop is large, then the executor automatically
splits the TPP into smaller TPPs.  The smaller TPPs use the {\tt
  CEXEC} instruction, conditioned on the hop number on the packet's
TPP header, to execute on a specific range of hops.}
\begin{eext}
\subsection{Deploying a TPP Application}\label{subsec:deploy}
Applications that use TPPs can be deployed in two ways: (i) standalone,
and (ii) piggy-backed.  Standalone apps use the raw TPP interface, and
have their own deployment strategy, because of their unique
requirements.  For example, RCP is a highly tuned implementation
within the OS kernel at every end-host.
\ignore{
\smallsec{Control-plane assisted apps}: The second class of applications
need access to a database of the network topology, switches, etc., to
achieve their functionality.  These applications are deployed on a
small subset of hosts in the network dedicated for this functionality,
on top of a control-plane framework (e.g. ONIX~\cite{koponen2010onix},
or Pyretic~\cite{monsanto2013composing}).
}

Piggy-backed applications are intrusive in
the sense that they attach TPPs to packets that flow through the
network.  There is a common pattern to such applications (e.g., network
troubleshooting and monitoring), which is abstracted away to make them
easy to deploy.  The programmer specifies the following:

\begin{itemize}[noitemsep,nolistsep,leftmargin=1em]
  \item Filter: An {\tt iptables} packet filter that specifies the subset of
    traffic through the network to which TPPs should be attached.  The
    filter also includes the sampling frequency and the application's
    priority.

  \item TPP: A compiled TPP (a string of bytes) that should be
    attached with packets.

  \item Aggregator: The aggregator is a per-node application that
    receives the application-specific, fully executed TPPs, and does
    post-processing.  For example, the OpenSketch monitoring
    application implements hash functions and summary data-structures.

  \item Collector: A software service that collects summaries from the
    aggregators spawned across the cluster.  The programmer specifies
    the service's virtual IP address; packets sent to this IP are load
    balanced across a replicated collector instances.
\end{itemize}

Once the programmer specifies the above inputs, a provisioning agent
creates a new application ID and verifies permissions by examining the
TPP\@.  The provisioning agent also spawns the aggregator and
collector which receive fully executed TPPs and does
application-specific processing.  Finally, the provisioning agent
configures appropriate end-host dataplane agents by invoking the {\tt
  add\_tpp} API call, completing the application setup.  Once the
application is set up, the collector will start receiving packets from
its aggregator agents on deployed hosts.
\end{eext}

\section{Implementation}
\label{sec:implementation}


We have implemented both hardware and software support needed for
TCPU: the distributed TCPU on the 10Gb/s NetFPGA platform, and a
software TCPU for the Open~vSwitch Linux kernel module.  The NetFPGA
hardware prototype has a four-stage pipeline at each port, with
64~kbit block RAM and 8~registers at each stage (i.e. a total of 1Mbit
RAM and 128 registers).  We were able to synthesize the hardware
modules at 160~MHz, capable of switching minimum sized (64Byte)
packets at a 40Gb/s total data rate.

The end-host stack is a relatively straightforward implementation: We
have implemented the TPP-CP, and the TPP executor (with support only
for the reliable and scatter-gather execution pattern) as Python
programs running in userspace.  The software dataplane is a kernel
module that acts as a shim between the network stack and the
underlying network device, where it can gain access to all network
packets on the transmit and receive path.  For filtering packets to
attach TPPs, we use {\tt iptables} to classify packets and tag them
with a TPP number, and the dataplane inserts the appropriate TPP by
modifying the packet in place.

\quad\section{Evaluation}
\label{sec:evaluation}
In \S\ref{sec:examples} we have already seen how TPPs enable many
dataplane applications.  We now delve into targeted benchmarks of the
performance of each component in the hardware and software stack.

\subsection{Hardware}
The cost of each instruction is dominated by the memory access
latency.  Instructions that only access registers complete in less
than 1~cycle.  On the NetFPGA, we use a single-port 128-bit wide block
RAM that has a read (or write) latency of 1~cycle.  We measured the
total per-stage latency by sending a hundreds of 4~instruction TPP
reading the clock from every stage, and found that the total per-stage
latency was exactly 2 cycles: thus, parsing, execution, and packet
rewrite all complete within a cycle, except for {\tt CSTORE}, which
takes 1 cycle to execute (excluding the time for accessing operands
from memory).

\begin{table}
\centering\small
\begin{tabular}[t]{|l|r|r|}\hline
{\bf Task}     & {\bf NetFPGA}    & {\bf ASICs} \\\hline
Parsing        &  $<1$ cycle      & 1 cycle  \\\hline
Memory access  &  1 cycle         & 2--5 cycles \\\hline
Instr. Exec.: {\tt CSTORE}   &  1 cycle         &   10 cycles      \\\hline
Instr. Exec.: (the rest)   &  $<1$ cycle      & 1 cycle  \\\hline
Packet rewrite &  $<1$ cycle      & 1 cycle  \\\hline
Total~per-stage&  2--3 cycles        & 50--100 cycles$\dagger$ \\\hline
\end{tabular}\caption{Summary of hardware latency costs.
$\dagger$The ASIC's per-stage cost is estimated from the total
end-to-end latency (200--500ns) and dividing it by the number of
stages (typically 4--5).  This does not include packetization latency,
which is another $\sim$50ns for a 64Byte packet at
10Gb/s.}\label{tab:latency}\vspace{-1em}
\end{table}

The latency cost is different in a real switch: From personal
communication with multiple ASIC designers~\cite{sarang-personal,barefoot-personal}, we learned that 1GHz ASIC chips in
the market typically use single-port SRAMs 32--128bits wide, and have a
2--5~cycle latency for every operation (read/write).  This means that
in the worst case, each load/store instruction adds a 5 cycle latency, and a {\tt
CSTORE} adds 10 cycles.  Thus, in the worst case, if every instruction
is a {\tt CSTORE}, a TPP can add a maximum of 50ns latency to the
pipeline; to avoid losing throughput due to pipeline stalls, we can
add 50ns worth of buffering (at 1Tb/s, this is 6.25kB for the entire
switch).  However, the real cost is likely to be smaller because the
ASIC already accesses memory locations that are likely to be accessed
by the TPP that is being executed: For instance, the ASIC always looks
up the flow entry, and updates queue sizes for memory accounting, so
those values needn't be read twice.

Though switch latency costs are different from that of the NetFPGA,
they do not significantly impact packet processing latency, as in a
typical workload, queueuing and propagation delays dominate end-to-end
latency and are orders of magnitude larger.  Even within a switch, the
unloaded ingress-egress latency for a commercial ASIC is about 500ns
per packet~\cite{asiclatency}.  The lowest-latency ASICs are in the
range of about 200ns per packet~\cite{fm4000}.  Thus, the extra 50ns
worst-case cost per packet adds at most 10--25\% extra latency to the
packet.  Table~\ref{tab:latency} summarizes the latency costs.

\begin{table}
\centering\small
\begin{tabular}[t]{|l|r|r|r|}\hline
{\bf Resource}  & {\bf Router} & {\bf +TCPU}  &  {\bf \%-extra}\\\hline
Slices          & 26.8K &  5.8K & 21.6\% \\\hline
Slice registers & 64.7K & 14.0K & 21.6\% \\\hline
LUTs            & 69.1K & 20.8K & 30.1\% \\\hline
LUT-flip flop pairs & 88.8K & 21.8K & 24.5\% \\\hline
\end{tabular}
\caption{Hardware cost of TPP modules at 4 pipelines in the NetFPGA
(4 outputs, excluding the DMA pipeline).}\vspace{-1em}
\label{tab:netfpga-cost}
\end{table}

\smallsec{Die Area}:  The NetFPGA costs are summarized in Table~\ref{tab:netfpga-cost}.
Compared to the single-stage reference router, the costs are within
30.1\% in terms of the number of gates.  However, gate counts by
themselves do not account for the total area cost, as logic only
accounts for a small fraction of the total area that is dominated by
memory.  To assess the area cost for a real switch, we use data from
Bosshart et al.\ ~\cite{glen2013hardware}.  In their paper, the
authors note that the extra area for a total of 7000 processing
units---which support instructions that are similar to the
TCPU---distributed across all match-action stages, accounts for less
than 7\% of the ASIC area~\cite[\S5.4]{glen2013hardware}.  We only
need $5\times{}64=320$ TCPUs, one per instruction per stage in the
ingress/egress pipelines; therefore, the area costs are not
substantial (0.32\%).


\subsection{End-host Stack}
The critical component in the end-host stack is the dataplane.  In the
transmit side, the dataplane processes every packet, matches against a
list of filters, and attaches TPPs.  We use a 4-core Intel core~i7
machine running Linux~3.12.6.

\begin{figure}[t]
\centering
\includegraphics[width=0.5\textwidth]{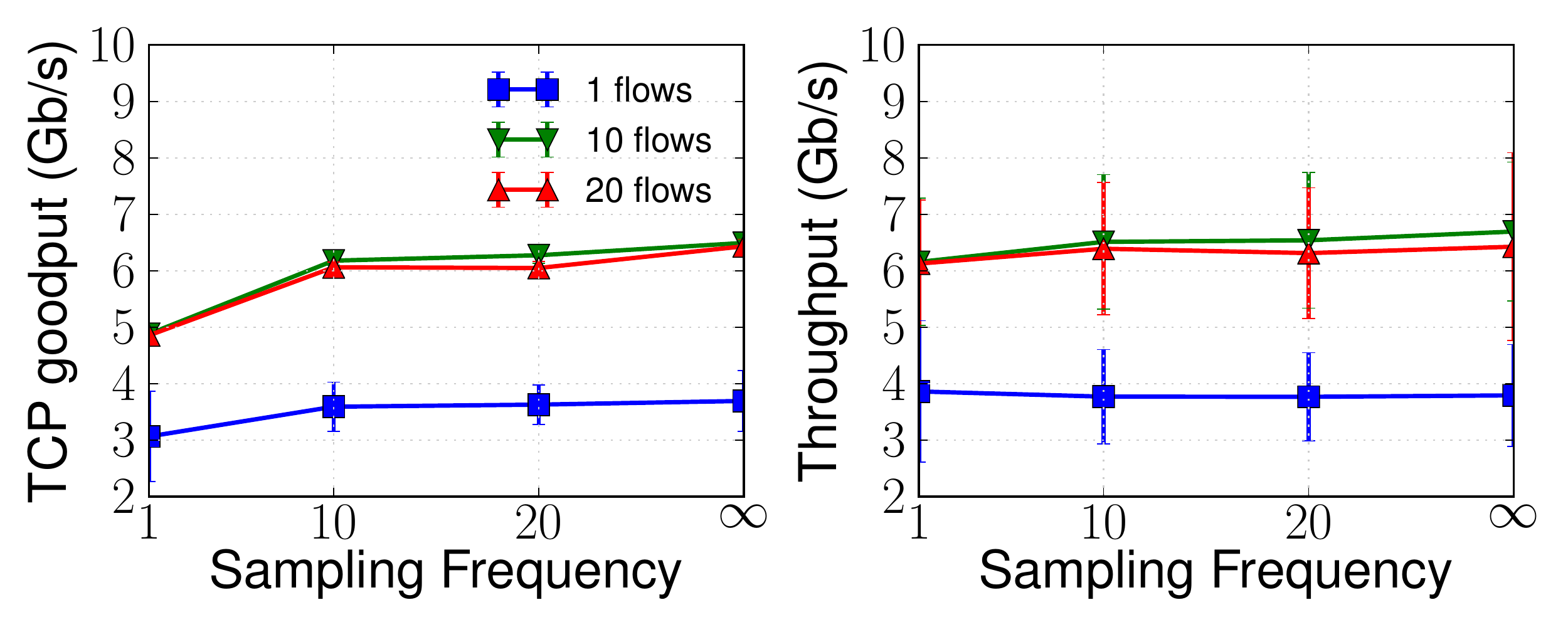}
\caption{Maximum attainable application-level and network throughput
  with a 260 byte TPPs inserted on a fraction of packets (1500Byte MTU
  and 1240Byte MSS).  A sampling
  frequency of $\infty$ depicts the baseline performance as no TPPs
  are installed.  Error bars denote the standard deviation.}\vspace{-1em}
\label{fig:throughput}
\end{figure}

Figure~\ref{fig:throughput} shows the baseline throughput of a single
TCP flow, without segmentation offloads, across a virtual ethernet
link, which was able to push about 4Gb/s traffic with one TCP flow,
and about 6.5Gb/s of traffic with 20 flows.  After
adding TPPs, the throughput of the TCP flow reduces, depending on the
(uniform random) sampling frequency.  If the sampling frequency is
infinite, none of the packets have TPPs, which denotes the best
possible performance in our setup.  As we can see, the network
throughput doesn't suffer much, which shows that the CPU overhead to
add/remove TPPs is minimal.  However, application throughput reduces
proportionally, due to header overheads.
\begin{table}\centering\small
\begin{tabular}{|l|r|r|r|r|r|}\hline
\multirow{2}{*}{\bf Match} & \multicolumn{5}{c|}{\bf \# Rules}  \\\cline{2-6}
 			& {\bf 0}   &  {\bf 1} &  {\bf 10} & {\bf 100} & {\bf 1000} \\\hline
	First   & 8.8 & 8.7 &  8.6 & 7.8 & 3.6  \\\hline
	Last    & 8.8 & 8.7 &  8.6 & 7.7 & 3.6  \\\hline
	All     & 8.8 & 8.7 &  8.3 & 6.7 & 1.4  \\\hline
\end{tabular}
\caption{Maximum attainable network throughput in Gb/s
with varying number of filters (1500Byte MTU).  The numbers
are the average of 5 runs.}\label{tab:filter-throughput}\vspace{-1em}
\end{table}
Table~\ref{tab:filter-throughput} shows the impact on the number of
filters in the dataplane, and its effect on network throughput, under
three different scenarios: (i) `first' means we create flows that
always match the first rule, (ii) `last' means flows always match the
last rule, and (iii) `all' means there is at least one flow that
matches each rule.  In `first' and `last,' there are 10 TCP flows.  In
`all,' there are as many flows as there are number of rules (with at
least 10 flows).  Each rule matches on a TCP destination port.  As we
can see, there is little loss in throughput up to 10 rules.  With more
rules, throughput does drop, but there is no difference between
matching on the first (best case) and last rule (worst case) in the
filter chain.  With 1000 flows, other overheads (context switches)
result in much lower throughput.

\section{Limitations}
\label{sec:limitations}
Though TPPs help in a wide variety of tasks that were discussed in
\S\ref{sec:examples}, they are not a panacea to implement any
arbitrary functionality due to two
reasons: (i) the restricted instruction set, and (ii) restricted programming
model in which end-hosts initiate tasks.  As we have not presented a formal theory of
``network tasks,'' the classification below is neither complete nor
mutually exclusive; it is only meant to be an illustration.

\smallsec{Tasks that require per-packet computation}: The read and
write instructions in
TPPs limit end-hosts to high throughput network \emph{updates}, but
not arbitrary network computation.  As an example, consider
the task of implementing an active queue management scheme such as
Stochastic Fair Queueing, static priority, dynamic priority queueing
(e.g. pFabric~\cite{alizadeh2013pfabric}), and fair queueing.  These
tasks require fine-grained control over per-packet transmit and drop
schedules, which is better realized using dedicated hardware or
FPGAs~\cite{sivaraman2013no}.  In a similar vein, TPPs are not
expressive enough to scan packets for specific signatures (e.g.,
payload analysis using deep packet inspection).  Such tasks
are better served by other approaches (e.g., middlebox software,
or custom packet processors).

\cam{\smallsec{Tasks that are event-driven}: In the examples we
discussed, all TPPs originate at end-hosts.  This limits
end-hosts from implementing tasks that require precisely timed notifications
whenever there is some state change within the network.  For instance, TPPs by
themselves cannot be used
to implement flow control mechanisms (e.g., priority flow control, or PFC~\cite{pfc}),
or reactive congestion notifications such as Quantized Congestion
Notification~\cite{pan2007qcn} and FastLane~\cite{zats2013fastlane}.  Such
tasks require the network to send special packets when the queue occupancy
reaches a certain threshold.  However, this isn't a show-stopper for TPPs,
as end-hosts can proactively inject TPPs on a subset of packets and be notified
quickly of network congestion.}


\section{Discussion}
\label{sec:discussion}
In \S\ref{sec:examples}, we showed how TPPs enable end-hosts to access
network state with low-latency, which can then act on this state to
achieve a certain functionality.  This is attractive as it enables
interesting functionality to be deployed
at software-development timescales.  We now discuss a number of
important concerns that we haven't covered.

\smallsec{Handling Device Heterogeneity}:  There are two issues here:
instruction encoding, and statistics addressing.  First, instructions
are unlikely to be implemented in an ASIC as hardwired logic, but
using microcodes, adding a layer of indirection for platform specific
designs.  Second, \cam{we recommend having two address spaces: (i) a standardized
address space where a majority of the important statistics are preloaded
at known locations,
such as those identified by the OpenFlow standard~\cite{of1.4},
and (ii) a platform-specific address space through which additional
statistics, specific to vendors and switch generations can be accessed.
For dealing with multiple
vendors, TPPs can support an indirect addressing scheme, so that the
the compiler can preload packet memory with platform specific
addresses.  For example, to load queue sizes from a
Broadcom ASIC at hop 1, and an Intel ASIC at hop 2, the compiler
generates the TPP below, loading the word from {\tt 0xff00} for
Broadcom, and {\tt 0xfe00} for Intel, obtained out-of-band.  For
safety, the entire TPP is wrapped around a {\tt CEXEC} as follows:}

{\small
\begin{verbatim}
  CEXEC [Switch:VendorID], [Packet:Hop[0]]
  LOAD [[Packet:Hop[1]], [Packet:Hop[1]]
PacketMemory:
  Hop1: $BroadcomVersionID, 0xff00 (* overwritten *)
  Hop2: $IntelVersionID, 0xfe00
\end{verbatim}
}

\noindent The TPP compiler can query the ASIC vendor IDs
from time to time and change the addresses if the devices at a
particular hop suddenly change.  However, indirect addressing
limits the extent to which a TPP can be statically analyzed.

\smallsec{MTU issues}:  Piggy-backed TPPs are attached to packets at the
edge of a network (end-host or a border router).  Thus, if the
incoming packet is already at the MTU size, there would be no room to
add a TPP.  This is fortunately not a big issue, as many switches
support MTUs up to 9000 bytes.  \cam{This is already being done
today in overlay networks to add headers for network virtualization~\cite{conga}.}\ignore{We could selectively
enable jumbo frames for packets with a TPP, while still treating
non-TPP packets as if the MTU was 1500 bytes.}


\extended{\smallsec{Offloading more functionality to the network}:  Many network
tasks we present in this paper requires cooperation from all end-hosts
to achieve a task.  It is worth asking if including every end-host is
worth the trouble.  We argue that it is: Many tasks such as congestion
control (e.g. TCP), and monitoring (e.g. SNAP~\cite{yu2011profiling}),
are already implemented in a way where every server takes part in the
task.  Network functionality can have a more informed path to a
hardware implementation starting with the end-hosts.

\smallsec{Active networks and end-to-end arguments}:  Active Networks was
criticized for its complexity and lack of `killer applications,' but
the end-to-end principle does not completely rule it
out~\cite{bhattacharjee1997active}.  The networks we build are larger,
more complex, and more critical than ever, as they form the backbone
of all compute within and between datacenters.  To manage a network at
large scale, applications must have an unprecendented visibility into
the dataplane at timescales orders of magnitude faster than with the
best tools we have today.  We do not claim that TPPs are ``novel,'' as
they fall under the broad category of active networks.  However, we
have ruthlessly tried to keep the interface between the end-hosts and
dataplane to the bare minimum to make interesting applications
feasible.  We believe TPPs strike a delicate balance between what is
possible in hardware at line rate, and sufficient flexibility that
end-hosts can implement dataplane tasks.}

\section{Related Work}
TPPs represent a point in the broad design space of programmable
networks, ranging from essentially arbitrary in-band programs as
formulated by Active Network
proposals~\cite{tennenhouse2002towards,schwartz1999smart}, to
switch-centric programmable dataplane
pipelines~\cite{fm6000,baden2010field,kohler2000click,glen2013hardware},
to controller-centric out-of-band proposals such as
OpenFlow~\cite{mckeown2008openflow} and Simple Network Management
Protocol (SNMP).  We do not claim that the TPP approach is a fundamentally novel
idea, as it is a specific realization of Active Networks.  However, we
have been ruthless in simplifying the interface between the end-hosts
and switches to a bare minimum.  We believe TPPs strike a delicate
balance between what is possible in switch hardware at line rate, and
what is sufficiently expressive for end-hosts to perform a variety of
useful tasks.

TPPs superficially resemble Sprocket, the assembly language in Smart
Packets~\cite{schwartz1999smart}.  However, Sprocket represents a far
more expressive point in the design space.  It allows loops and larger
programs that would be hard to realize in hardware at line rate.  By
contrast, a TPP is a straight-line program whose execution latency is
deterministic, small, and known at compile time.  TPPs fully execute
on the fast-path (i.e., router ASIC), whereas Sprocket exercises the
slow-path (router CPU), which has orders of magnitude lower
bandwidth. \cam{TPPs also resemble the read/write in-band control
mechanism for ATM networks as described in a
patent~\cite{inband-patent}; however, we also focus extensively on how
to refactor useful dataplane tasks, and a security policy to safeguard
the network against malicious TPPs.  Wolf et al.~\cite{wolf2001design}
focus on designing a high performance Active Network router that
supports general purpose instructions.  It is unclear whether their
model allows end-hosts to obtain a consistent view of network state.
Moreover, it is unlikely that ASICs can take on general purpose
computations at today's switching capacities at a reasonable cost.
Furthermore, out-of-band control mechanisms such as OpenFlow and
Simple Network Management Protocol (SNMP) neither meet the performance
requirements for dataplane tasks, nor provide a packet-consistent
view of network state.}


There have been numerous efforts to expose switch statistics through
the dataplane, particularly to improve congestion management and
network monitoring.  One example is Explicit Congestion Notification
in which a router stamps a bit in the IP header whenever the egress
queue occupancy exceeds a configurable threshold.  Another example is
IP Record Route, an IP option that enables routers to insert the
interface IP address on the packet.  \cam{Yet another example is Cisco's
Embedded Logic Analyzer Module (ELAM)~\cite{elam} that traces the packet's
path inside the ASIC at layer~2 and layer~3 stages, and generates a
summary to the network control plane.}  Instead of anticipating future
requirements and designing specific solutions, we adopt a more
generic, protocol-independent approach to accessing switch state.

\section{Conclusion}
\ignore{Our goal in this paper is to enable end-hosts to flexibly measure and
control network behavior.  Our key insight is that providing end-hosts
with low-latency access to shared network state helps in realizing
this goal.  To maintain flexibility, we designed a programmable
interface to access network state.  To achieve low-latency, we
proposed tiny packet programs that query and modify network state
directly in the dataplane.  By identifying simple instructions that
can execute at line-rate in hardware, we were able to push the
responsibility of carrying out complex computation on network state to
end-hosts.  Our design to split functionality between end-hosts and
the network is largely driven by the key requirement to meet the
stringent performance requirements of the ASIC.}

We set out with a goal to rapidly introduce new dataplane
functionality into the network.  We showed how, by presenting a
programmatic interface, using which end-hosts can query and manipulate network state directly
using tiny packet programs.  TPPs support both a distributed
programming model in which every end-host participates in a task
(e.g., RCP* congestion control), and a logically centralized model
in which a central controller can monitor and program the network.
We demonstrated that TPPs enable a whole
new breed of useful applications at end-hosts: ones that can
work \emph{with} the network, have unprecedented visibility nearly
instantly, with the ability to tie dataplane events to \emph{actual}
packets, umambiguously isolate performance issues, and act on network
view without being limited by the control plane's ability to provide
such state in a timely manner.


\section*{Acknowledgments}
Vimalkumar thanks Brandon Heller, Kok-Kiong Yap, Sarang
Dharmapurikar, Srinivas Narayana, Vivek Seshadri, Yiannis Yiakoumis, Patrick Bosshart,
Glen Gibb, Swarun Kumar, Lavanya Jose, Michael Chan, Nick McKeown, Balaji Prabhakar,
and Navindra Yadav for helpful
feedback and discussions that shaped this work.  The authors also thank our shepherd
John Wroclawski and the anonymous SIGCOMM reviewers for their thoughtful reviews.

The work at Stanford was funded by
NSF FIA award CNS--1040190.  Opinions, findings, and conclusions do
not necessarily reflect the views of NSF or other sponsors.



\printbibliography

\extended{
\newpage\onecolumn
\appendix
\section*{Memory Map}\label{sec:memorymap}
In this section, we list the memory map to access switch statistics
and per-packet metadata that is usually tracked by
a switch.  This list of statistics and metadata is by no means complete,
nor necessarily tracked by all switches, but it serves as a reference list
of statistics we found useful when implementing new dataplane tasks.

\subsection*{Counters and statistics}\label{sec:stats}
We borrow heavily from the list of counters from the
OpenFlow~1.4 specification~\cite[Table~5]{of1.4}.
We introduce more useful counters
which would help monitoring and debugging the network.  As switch
vendors add more statistics, they could be made accessible to
TPPs using a vendor-specific memory map.  The vendor can then provide
documentation on using the memory map using a data sheet.

When we refer to some values as `stats,' or a `stats block' we track
the following four counters: packets, bytes, rate of packets, and rate
of bytes.  For example, the `Lookup stats' block tracks the aggregate
number of bytes and packets that resulted in a lookup on the table.
This is useful to know for every table, as some packets (e.g., an
ARP packet) may not match all tables (e.g., the layer 3 routing table).

\begin{table}[h]\centering
\begin{tabular}{|l|p{0.5\textwidth}|}\hline
{\bf Value Name}      & {\bf Purpose} \\\hline
\multicolumn{2}{c}{Per ASIC.  Namespace {\tt [Switch:]}} \\\hline
Switch ID             & A unique ID given to a switch \\\hline
Version Number        & A global counter that tracks the generation of 
                        the switch forwarding state \\\hline
Clock/uptime          & The duration, in clock cycles, for which the switch
						has been online \\\hline
Clock frequency       & The clock frequency of the switching ASIC in
						cycles per second \\\hline

\multicolumn{2}{c}{Per Flow Table.  Namespace {\tt [Stage\$i:]} for the $i^{\rm th}$ stage.} \\\hline
Version Number        & A per flow table version number that monotonically
						increases on every flow update \\\hline
Reference Count       & Number of active flow entries in the table \\\hline
Lookup stats          & Number of packets and bytes that resulted in a lookup in the table \\\hline
Match stats           & Number of packets and bytes that matched some entry in the table \\\hline

\multicolumn{2}{c}{Per Flow Entry.  Namespace {\tt [FlowEntry\$i:]} for the $i^{\rm th}$ stage.} \\\hline
Insert clock          & The clock cycle when the flow entry was installed in the table \\\hline
Match stats           & Number of packets and bytes that resulted in a match on this entry \\\hline

\multicolumn{2}{c}{Per Port.  Namespace {\tt [Link\$i:]} for the $i^{\rm th}$ link.} \\\hline
Queued stats          & Number of packets and bytes waiting on this port to be transmitted \\\hline
Transmit stats        & Stats block for packets transmitted on this port \\\hline
Receive stats         & Stats block for packets received on this port \\\hline
Drop stats            & Stats block for drops on this port \\\hline
Error stats           & Stats block for CRC/other bit errors on this port \\\hline
Port status           & Status bits (up/down/maintenance/etc.) for this port \\\hline

\multicolumn{2}{c}{Per Queue.  Namespace {\tt [Queue\$i\$j:]} for the $j^{\rm th}$ queue on the $i^{\rm th}$ link.} \\\hline
Scheduling configuration block & This memory block contains counters and statistics
                                 that pertain to the scheduling algorithm (e.g., deficit
                                 round robin weight, quantum, etc.). \\\hline

Queued stats          & Stats block for packets currently queued \\\hline
Transmit stats        & Stats block for transmitted packets \\\hline 
Receive stats         & Stats block for received packets \\\hline
Drop stats            & Stats block for dropped packets \\\hline

\end{tabular}\caption{Statistics required from the ASIC, based on
the standard values tracked by an OpenFlow~1.4 capable switch.  These
statistics must be accessible to the TPP.}
\end{table}

\newpage
\subsection*{Per-packet metadata}\label{sec:perpacket}
Recall that the packet is processed once at the ingress pipeline,
where its output port(s) is (are) determined.  We expect the
statistics listed in Table~\ref{tab:ingress-stats} to be available to every TPP.
These statistics are not shared across packets.  It is analogous to
the {\tt /proc/self} interface that a process can use to access
its own statistics such as memory usage.  Similarly, Table~\ref{tab:egress-stats}
lists statistics that should be available to every TPP at the egress.

\begin{table}[t]\centering
\begin{tabular}{|l|p{0.5\textwidth}|}\hline
{\bf Value Name}      & {\bf Purpose} \\\hline
\multicolumn{2}{c}{Namespace: {\tt [PacketMetadata:]}.} \\\hline

Input port  & The input port number on which the packet
		arrived to the switch \\\hline

Input port statistics &  An indirection to the stats
		block tracking the port's counters \\\hline

Output port bitmap  & A bitmap indicating the port(s) out
		of which the packet will be forwarded.  The bitmap is initialized
		to 0.  If, at the end of the ingress pipeline, the bitmap is
		still set to 0, then the packet will be dropped.
		\\\hline

Matched flow entry  & An index into the flow entry that
		the packet matches at each table \\\hline

Matched flow entry stats &  Each flow entry tracks
		various statistics as shown above.  This is an indirection to
		the stats block for the flow entry counters. \\\hline

Enqueued queue ID &  An index into the queue at the output
		port into which the packet is queued for transmission \\\hline

Enqueued queue's stats &  An indirection to the stats
		block tracking the queues's counters \\\hline

Packet fields &  All packet fields that were parsed by the parser
		for this packet.  This includes all the standard OpenFlow~1.4~\cite{of1.4}
		fields, and our proposed fields for the TPP~(\S\ref{subsec:tppformat}).
		\\\hline
\end{tabular}
\caption{Per-packet metadata that must be available to every TPP at the ingress pipeline.}
\label{tab:ingress-stats}
\end{table}


\begin{table}[t]\centering
\begin{tabular}{|l|p{0.5\textwidth}|}\hline
{\bf Value Name}      & {\bf Purpose} \\\hline
\multicolumn{2}{c}{Namespace: {\tt [PacketMetadata:]}.} \\\hline
Output port  &  The port through which the packet is currently
		being forwarded out of \\\hline

Output port statistics &  An indirection to the stats
		block tracking the port's counters \\\hline

Output queue  & The queue on which the packet was scheduled \\\hline

Output queue statistics & An indirection to the stats
		block tracking the queue's counters \\\hline
\end{tabular}
\caption{Per-packet metadata that must be available to every TPP at the egress pipeline.}
\label{tab:egress-stats}
\end{table}


}

\end{sloppypar}
\end{document}